\documentclass[journal]{IEEEtran}
\usepackage{cite}	
\ifx\pdfoutput\undefined
\usepackage{graphicx}
\else
\usepackage[pdftex]{graphicx}
\fi
\usepackage{epsfig} 
\usepackage{psfrag}    
\usepackage{subfigure} 
\usepackage{url}       
\usepackage{amsmath}
\usepackage{array}
\usepackage{multirow}
\usepackage{threeparttable}
\usepackage{units}
\usepackage{booktabs}
\usepackage{xcolor}
\usepackage{amsmath,amsfonts,amssymb}
\usepackage{float}
\usepackage{footnote}
\ifx\pdfoutput\undefined
\fi
\usepackage{threeparttable}
\usepackage{epstopdf}

\begin{document}

\title{Cryo-CMOS Band-gap Reference Circuits for Quantum Computing}

\author{Yuanyuan~Yang, 
				Kushal~Das, 
				Alireza~Moini,
				and~David. J. Reilly

\thanks{Y. Yang is with ARC Centre of Excellence for Engineered Quantum Systems, School of Physics, the University of Sydney, Sydney, NSW 2006, Australia.}
\thanks{K. Das and A. Moini are with Microsoft Quantum Sydney, the University of Sydney, Sydney, NSW 2006, Australia.}
\thanks{D. J. Reilly is with ARC Centre of Excellence for Engineered Quantum Systems, School of Physics, the University of Sydney, Sydney, NSW 2006, Australia and he is also with Microsoft Quantum Sydney, the University of Sydney, Sydney, NSW 2006, Australia.}
}

\maketitle

\begin{abstract}

The control interface of a large-scale quantum computer will likely require electronic sub-systems that operate in close proximity to the qubits, at deep cryogenic temperatures. Here, we report the low-temperature performance of custom cryo-CMOS band-gap reference circuits designed to provide stable voltages and currents on-chip, independent of local temperature fluctuations. Our circuits are fabricated in \unit[0.35]{$\boldsymbol\mu$m} Silicon Germanium (SiGe) BiCMOS and \unit[28]{nm} Fully Depleted Silicon On Insulator (FDSOI) CMOS processes, and we compare the performance of each. Beyond their specific application as low-power references, these circuits are ideal test-vehicles for developing design approaches that mitigate the adverse effects of cryogenic temperatures on circuit performance.

\end{abstract}

\begin{IEEEkeywords}

BGR, voltage reference, current reference, SiGe, BiCMOS, cryogenic temperature, cryogenic electronics, cryogenic CMOS, cryogenic ASIC, quantum computer.

\end{IEEEkeywords}

\IEEEpeerreviewmaketitle

\section{Introduction}

\IEEEPARstart{R}{eference} circuits producing stable and precise voltages or currents are considered foundational elements of modern integrated circuit (IC) design, and in general, set the limit on the performance of many analog systems. For semiconductor devices, it was realized in the 1950s that the material system itself readily presents a natural voltage reference in the form of the energy band-gap between valence and conduction bands, typically of order \unit[1]{eV} or so~\cite{shockley}. Thermally activated currents however, depend exponentially on the ratio of the thermal energy  to the bandgap, $E_\text{G}$, rendering a direct approach strongly susceptible to temperature fluctuations. The solution, known now for many decades~\cite{Semiconductor-Voltage-Standard-1964}, is in the design of a circuit that first generates a current proportional to temperature and then uses this current to produce a voltage that can cancel the original temperature dependent signal (to first order). These so-called band-gap references (BGRs) are now ubiquitous circuit blocks underpinning applications that span stand-alone voltage regulators, complex mixed-signal ICs, and system on chips (SoCs).

Although the purpose of a BGR is to produce a stable voltage that is largely immune to temperature variations, the basic semiconductor physics underpinning the functionality of the transistors themselves can limit the circuit to applications operating near room temperature. This presents a challenge for emerging technologies such as medical detectors~\cite{Cryo-Medical-2018,Cryo-NMR-DiffAmp-2018}, space applications~\cite{Cryo-DC2DC-2003,LAr-ASIC-2011,ESA-LNA-2012,CESAR-2014,Cryo-Ka-Rx-2018} and quantum computers~\cite{Cryo-Arch1,Cryo-FPGA,Delft-Cryo-CMOS,Google-TRx}, which will likely require complex control interfaces operating at deep-cryogenic temperatures. Under these conditions new phenomena emerge in the behavior of transistors including carrier freeze-out~\cite{Cryo-bulk-1986,Freeze-out-1989,Quasi-CV-1995,LDMOS-Freezeout-2010}, kink-effect~\cite{Cryo-bulk-1986,Analysis-kink-1990,Processing-step-kink-2001} and threshold voltage shift~\cite{Temp-dependent-1968,Cryo-bulk-1986,Extreme-temp-electronics-2003,VTH-10nmSOI-2004,Temp-effect-SOI-MOS-2004,28nm-FDSOI-RO-2018,Cryo-behavior-2018}. At the same time, cryogenic operation also requires inherently low-power operation and noise specifications near fundamental limits.

\begin{figure*}[!t]
\centering
\vspace*{-0.2cm}
\hspace*{-0cm}
\includegraphics[width=0.7\textwidth, angle=0]{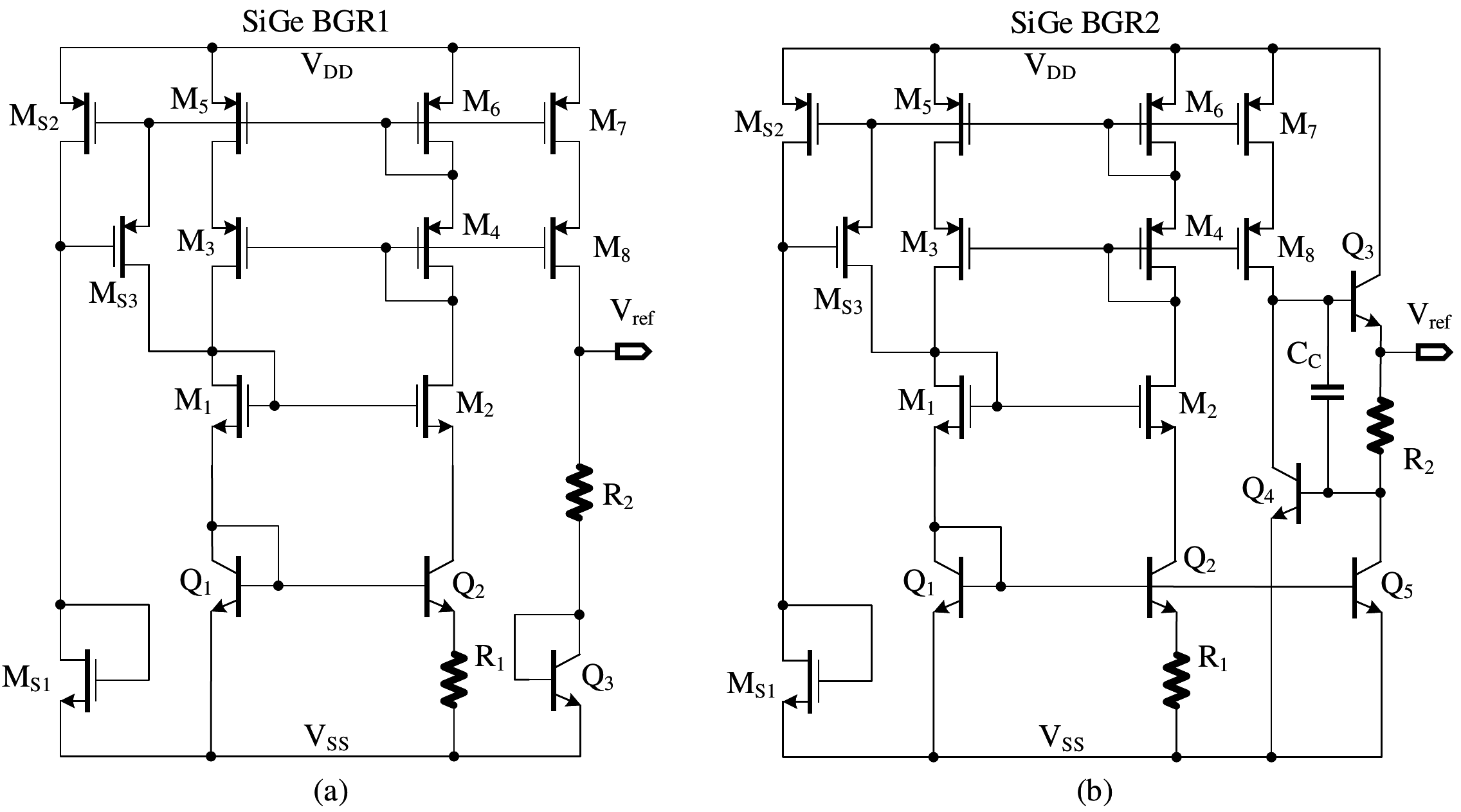}
\vspace*{-0.2cm}
\caption{BGR designs in \unit[0.35]{$\mu$m} SiGe process: (a), conventional current mirror BGR topology, we name it SiGe BGR1; (b), modified SiGe BGR from~\cite{sub-1K-BGR-2009}, we name it SiGe BGR2. All the BJTs in the reference circuits are SiGe HBTs. In BGR1, $\text{Q}_\text{1}$, $\text{Q}_\text{2}$ and $\text{Q}_\text{3}$ are implemented in a 5 by 7 unit cells array. $\text{Q}_\text{1}$, $\text{Q}_\text{2}$ has a ratio of 1 : 32 and $\text{Q}_\text{3}$ is implemented with the rest of the 2 unit cells. In BGR2, $\text{Q}_\text{1}$ - $\text{Q}_\text{5}$ are arranged in a 6 by 6 unit cells array, with $\text{Q}_\text{1}$, $\text{Q}_\text{2}$ in a ratio of 1 : 32 and $\text{Q}_\text{3}$ - $\text{Q}_\text{5}$ are implemented with 1 unit cell each.}
\vspace*{-0cm}
\label{fig:BGRs_SiGe}
\end{figure*}

Our focus on cryogenic reference circuits is motivated firstly by the need for stable, low-power voltage and current sources on-chip, dynamically configurable and tightly integrated with auxiliary qubit devices in a scaled-up control system. In particular, for CMOS-based control systems that can generate significant heat relative to the available cooling power at cryogenic temperatures, the use of temperature-stable integrated reference circuits can help mitigate the adverse effects arising from on-chip temperature variations. A secondary motivation stems from the use of these circuits as generic test-vehicles for examining circuit design techniques and general approaches to nullify the unwanted effects of the cryogenic environment. Bandgap references are well suited for this purpose since, at their core, they employ feedback to compensate for undesirable temperature dependence. Beyond voltage references, similar feedback approaches can be used to establish a tool-kit of circuit techniques for use in architecting cryogenic control systems.

Here we demonstrate cryogenic operation of voltage and current reference circuits specifically designed to enable applications in quantum computing, where classical electronic sub-systems are required to interface with quantum devices. Extending the use of bandgap references to the deep cryogenic regime, we compare the low-temperature performance of circuits fabricated in \unit[0.35]{$\mu$m} SiGe BiCMOS and \unit[28]{nm} Fully Depleted Silicon On Insulator (FDSOI) CMOS processes. By way of a simple model, for each circuit we undertake a detailed analysis of how variations in temperature and fabrication-dependent parameters lead to uncertainty in the output of the reference. 

This paper is organized as follows: Section II presents our choice of technologies and circuit topologies of the designs; Section III presents experimental results of our reference circuits measured in cryogenic probe stations and dilution refrigerators over a wide temperature range; Section IV is the conclusion.

\section{Process and design choices}

The original~\cite{BGR1971} and modified~\cite{BGR1974} BGR topologies have been utilized as one of the essential building blocks in analog ASICs. When it comes to deep cryogenic temperatures, silicon-based bipolar junction transistors (BJTs), which rely on thermally-ionised dopants, typically freeze-out to the extent that conventional BGRs are rendered non-operational. The SiGe hetero-junction bipolar transistor (HBT), due to its engineered band structure, does not suffer from carrier freeze-out~\cite{SiGe-extreme-temp-Cressler-2005,0.8T-SiGe-HBT-Cressler-2014,SiGe-HBT-70mK-Cressler-2017,RF-LNA-SiGe-HBT-Cressler-2018} to the same extent and can be employed to design a BGR voltage reference. 

MOS-based voltage references~\cite{DTMOSTs-1999,Mutual-VT-U-2001,Sub1V-Nanopower-2007,300nW-15ppm-2009,Mutual-VTH-2012,500nA-CMOS-Only-2014,Nano-Watt-MOS-Only-2018} are another alternative for cryogenic applications, despite being susceptible to an increase in uncertainty of threshold voltage that stems from process variations. This type of reference voltage is usually generated based on the diode connected MOS transistor. MOS devices are known to be functional at cryogenic temperatures~\cite{Cryo-28nm-bulk-2017,Mismatch-CMOS-2018,Bulk-FDSOI-28nm-2018} as the channel formation based on inversion does not require thermally activated carriers.  Nevertheless, there are two significant challenges to implement MOS-only voltage references. The first one is commonly referred to as ``kink effect" in bulk CMOS processes~\cite{Cryo-bulk-1986,Analysis-kink-1990,Processing-step-kink-2001}, in which the increase in resistance of the transistor body leads to a gating effect as charge becomes unable to dissipate. This phenomena may increase the uncertainty of a node voltage, reduce the gain of a control circuit and alter the stability margin of a feedback loop. The second challenge is the significant threshold voltage $V_\text{TH}$ variation that occurs as bulk MOS devices are cooled from room  to cryogenic temperatures~\cite{Temp-dependent-1968,Cryo-bulk-1986,Extreme-temp-electronics-2003,Cryo-behavior-2018}. Addressing the challenges of bulk CMOS, we additionally implement reference circuits in a \unit[28]{nm} FDSOI process. The fully-depleted feature of this process eliminates the unwanted ``kink effect"~\cite{Kink-FDSOI-2017}. Meanwhile, this technology offers back-gate control nodes to both PMOS and NMOS transistors~\cite{28nm-FDSOI-Quantum-Computer-2017}, enabling in situ control of threshold voltage at cryogenic temperature.

\begin{figure*}[!t]
\centering
\vspace*{-0.2cm}
\hspace*{-0cm}
\includegraphics[width=0.9\textwidth, angle=0]{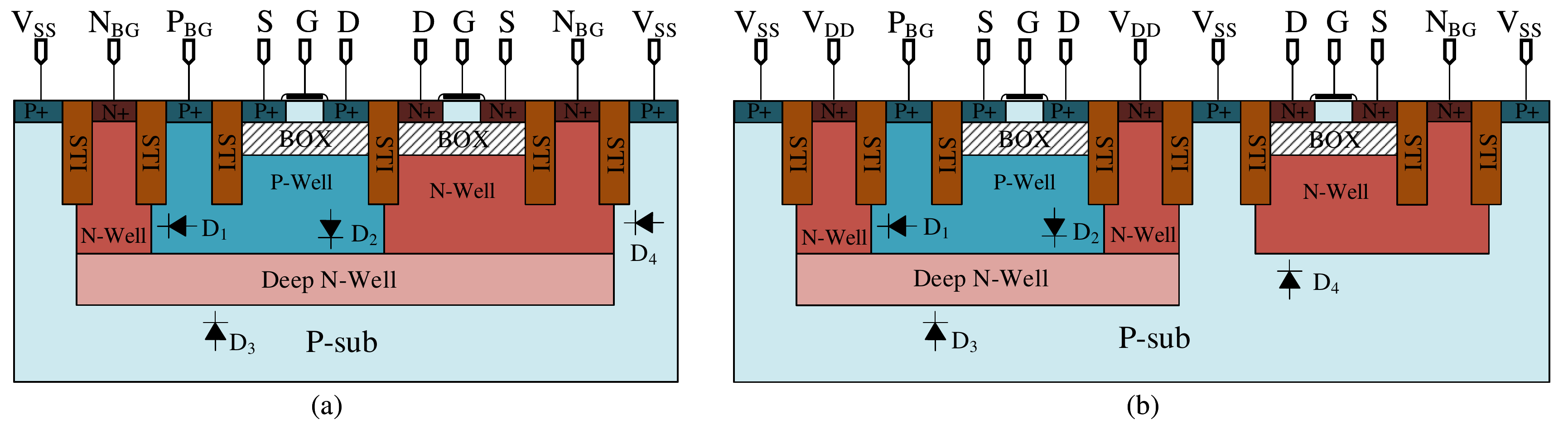}
\vspace*{-0.2cm}
\caption{Cross section view of the FDSOI technology. $\text{N}_\text{BG}$ is the back-gate control terminal of NMOS transistors and $\text{P}_\text{BG}$ is the back-gate control terminal of PMOS transistors.}
\vspace*{-0cm}
\label{fig:cross_section}
\end{figure*}

\subsection{BGR Designs in \unit[0.35]{$\mu$m} SiGe Process}

The first generation of our low-temperature reference circuits are designed and fabricated in \unit[0.35]{$\mu$m} SiGe BiCMOS process. The process provides CMOS and HBT devices on the same die. We implement two versions of current mirror based SiGe BGRs, as shown in Fig. \ref{fig:BGRs_SiGe}. The circuit on the left hand side of the figure, SiGe BGR1, follows a conventional topology of the BGR design with the purpose of benchmarking cryogenic performance. The design in Fig. \ref{fig:BGRs_SiGe}(b), SiGe BGR2, is similar to that reported in~\cite{sub-1K-BGR-2009}. It uses a modified start-up circuit and a frequency compensated output stage. For some of our applications, the output stage needs to directly drive other electronic systems at a different temperature via a cable. This introduces a capacitive load, and together with any parasitic capacitance at the base terminal of $\text{Q}_\text{4}$, will reduce the stability margin of the output stage. The compensation capacitor $\text{C}_\text{C}$ in Fig. \ref{fig:BGRs_SiGe}(b) provides a fast signal coupling path which enhances the stability.

The SiGe BGRs in Fig. \ref{fig:BGRs_SiGe} employ first-order temperature co-efficient (TC) cancellation at a targeted temperature $T_\text{0}$. For SiGe BGR2, the output voltage is formed by a complementary to absolute temperature (CTAT) voltage at the base terminal of $\text{Q}_\text{4}$ plus the proportional to absolute temperature (PTAT) voltage drop on $\text{R}_\text{2}$. Without exploiting detailed low-temperature models, it is neither straightforward to directly set $T_\text{0}$ to a target cryogenic temperature (\unit[4]{K}) nor possible to predict the reference voltage uncertainties due to (process-voltage-temperature) PVT variations. Nevertheless, a basic analysis of the circuit behavior provides insight into how these circuits function at low temperatures and how the room-temperature non-idealities translate into cryogenic-temperature non-idealities.

The output voltage variations due to BGR internal error sources are studied for a room-temperature BGR design~\cite{bgr-pvt-2005}. Particularly, in conventional BGR designs, the device mismatches are the major error contributors. At cryogenic temperatures, it has been reported~\cite{KD-mismatch-2014,Mismatch-CMOS-2018} that the threshold mismatch in CMOS current mirrors becomes larger than its room-temperature value. We can readily identify $\text{M}_\text{5}$ - $\text{M}_\text{7}$, $\text{R}_\text{1}$ and $\text{R}_\text{2}$ are matching critical devices in BGR1, see Fig. \ref{fig:BGRs_SiGe}(a). As the current mirrors set the PTAT current and output current value; the resistors set the PTAT part of the output voltage. In BGR2, see Fig. \ref{fig:BGRs_SiGe}(b), in addition to that of BGR1, the BJTs $\text{Q}_\text{1}$, $\text{Q}_\text{2}$ and $\text{Q}_\text{5}$ are match critical elements, as $\text{Q}_\text{5}$ mirrors the PTAT current in the output stage.

\subsection{Reference Designs in \unit[28]{nm} FDSOI}

\begin{figure}
\centering
\vspace*{-0cm}
\hspace*{-0.2cm}
\includegraphics[width=0.5\textwidth, angle=0]{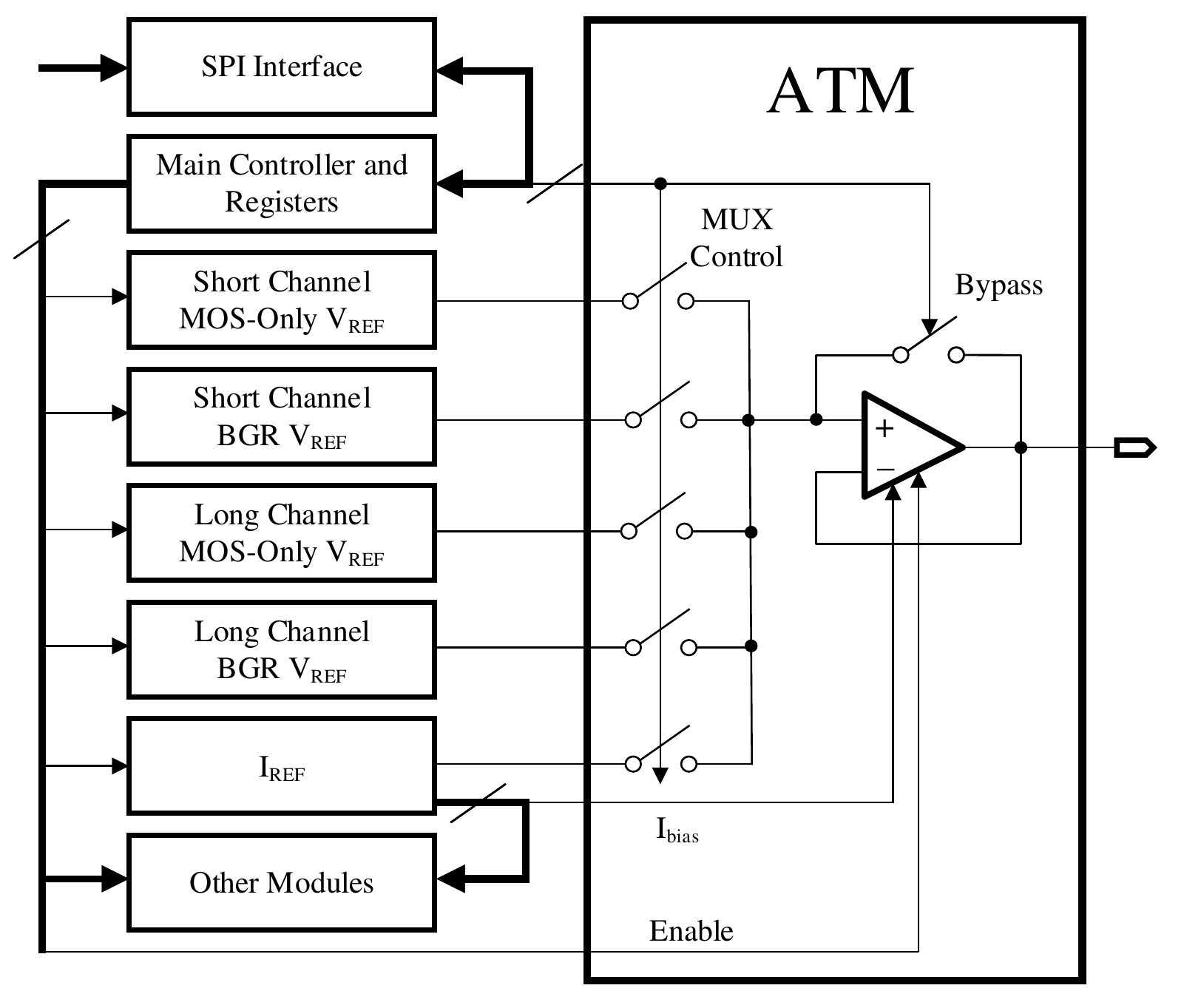}
\vspace*{-0.2cm}
\caption{Simplified test architecture of our \unit[28]{nm} FDSOI CMOS SoC. This architecture is modified according to the scope of this paper.}
\vspace*{-0cm}
\label{fig:GB_Arch}
\end{figure}

\begin{figure}
\centering
\vspace*{-0cm}
\hspace*{-0.2cm}
\includegraphics[width=0.5\textwidth, angle=0]{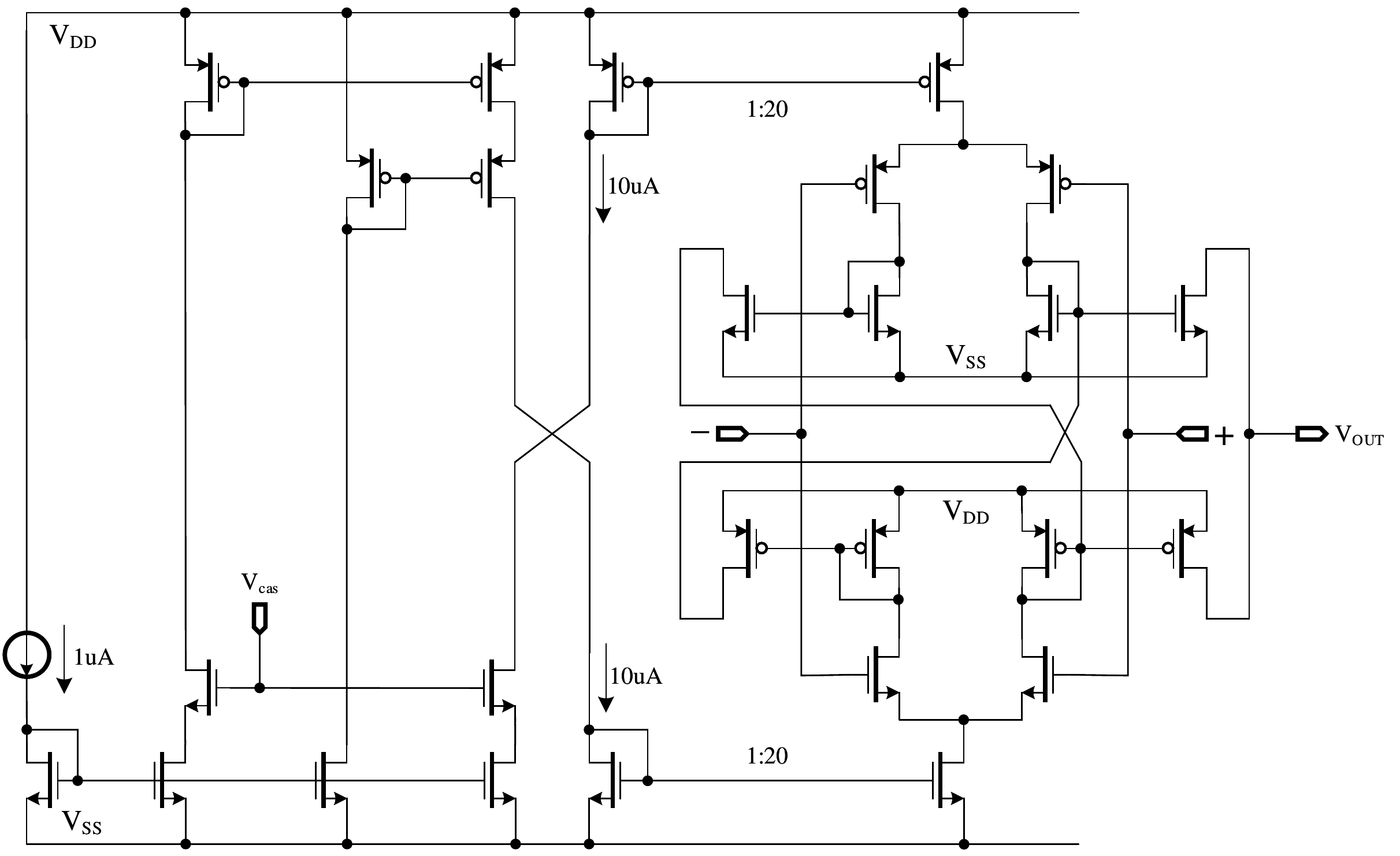}
\vspace*{-0.2cm}
\caption{Circuit schematic diagram of the ATM buffer amplifier.}
\vspace*{-0cm}
\label{fig:ATM_BUF}
\end{figure}

Despite the potential suitability of our SiGe BGRs for cryogenic operation, the use of this technology platform for complex, mixed-signal applications is limited, in particular with respect to ultra-low power operation required for quantum computing. The \unit[28]{nm} FDSOI CMOS technology provides a platform for implementing logic and analog circuits using a \unit[1]{V} supply to achieve better power efficiency. There has also been work showing sub-\unit[1]{V} operations of low-voltage BGR circuits~\cite{Sub-1V-BGR-1999,Sub-1V-BGR-2001,Sub-1V-BGR-2002,Sub-1V-BGR-2007,Sub-1V-BGR-2019}. At low temperatures, the increase in threshold voltages for both PMOS and NMOS transistors reduces the voltage headroom for proper circuit operations, which pose significant design challenges for analog circuits. For this reason, we implement our reference circuits using I/O devices with thicker oxide and operate them at \unit[1.8]{V} power supply.

\begin{figure*}[!t]
\centering
\vspace*{-0.2cm}
\hspace*{-0cm}
\includegraphics[width=0.8\textwidth, angle=0]{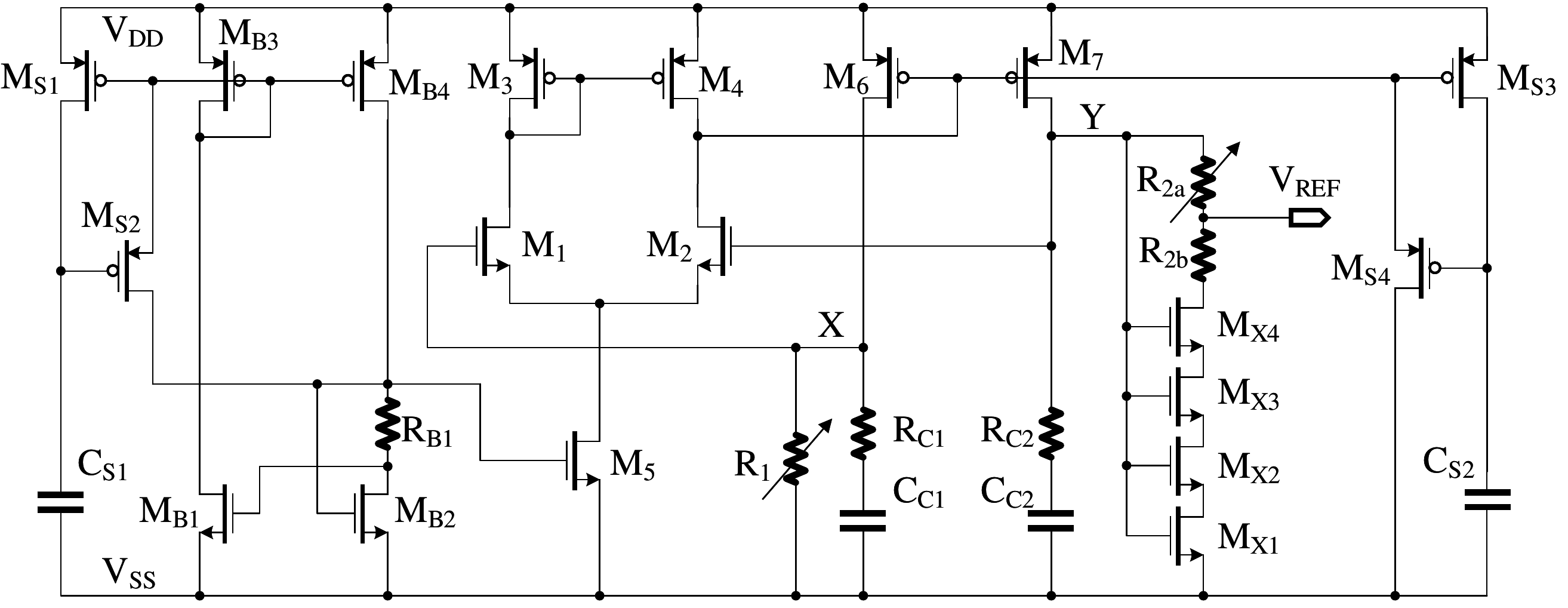}
\vspace*{-0.2cm}
\caption{MOS-only voltage reference implemented using \unit[1.8]{V} I/O transistors in \unit[28]{nm} FDSOI CMOS process.}
\vspace*{-0cm}
\label{fig:BGR_ZTC}
\end{figure*}

There are two different strategies to configuring the back-gate control terminals in this technology, as shown in Fig. \ref{fig:cross_section}. From both of the cross sections in the figure, we can see that the NMOS transistor's back-gate control voltage $\text{N}_\text{BG}$ cannot go below the ground potential $\text{V}_\text{SS}$. This means we cannot obtain high threshold voltage NMOS devices via back-gate controls. In fact, we would like to reduce the threshold voltages of the MOS transistor for our low-temperature applications. In the approach shown on the left in this figure, we can see that the PMOS back-gate control voltage $\text{P}_\text{BG}$ cannot be higher than the NMOS back-gate voltage $\text{N}_\text{BG}$ and when it comes to extreme high back-gate biases (beyond the value of power supply voltages), this scheme will have higher voltage stress at the PN junction interface of $\text{D}_\text{1}$ and $\text{D}_\text{2}$ compared to the cross section shown on the right of the figure. The scheme on the right mimics a traditional triple well CMOS technology, and it provides more freedom for the choice of back-gate voltages due to the physical separation of the $\text{P}_\text{BG}$ and $\text{N}_\text{BG}$. In Fig. \ref{fig:cross_section} using a clean power supply, $\text{V}_\text{DD}$, better noise isolation between the PMOS and NMOS devices can be achieved, as they are physically isolated. For the above reasons, we place and route our digital blocks using the cross section shown in Fig. \ref{fig:cross_section}(a), as it is more compact in layout. Reference blocks are implemented using the scheme shown in Fig. \ref{fig:cross_section}(b).

Our reference voltage and current generation circuits have been implemented as part of an SoC. The simplified chip architecture is presented in Fig. \ref{fig:GB_Arch}. In this chip, commands are sent through a serial-parallel-interface (SPI). An Analog Test MUX (ATM) is implemented to buffer the voltage reference signals when driving the load from external test and measurement instruments. The buffer uses a rail-to-rail amplifier topology and its circuit diagram is presented in Fig. \ref{fig:ATM_BUF}. The ATM can be bypassed when performing current reference tests.

We implement the reference circuits by configuring transistors of different length. In the short channel version of the circuit, the long transistors ($L > $ \unit[0.5]{$\mu$m}) are replaced with a number of shorter series transistors with $L < $ \unit[0.5]{$\mu$m}  The current reference circuits are all implemented using short channel devices.

The MOS-only voltage reference shown in Fig. \ref{fig:BGR_ZTC} is based on a modified form described in~\cite{BGR_ZTC}. We have added two dynamic start-up circuits, because at low temperatures the start-up mechanism which relies on leakage current may not function properly. Without a ``kick'' on the gate of $\text{M}_\text{6}$ and $\text{M}_\text{7}$, the node voltage on X and Y will be almost zero as there is no current flowing through both of the two branches. At room temperature, there may be a small leakage current flowing through $\text{M}_\text{6}$ and $\text{M}_\text{7}$. The voltage at node X will still be around zero and the voltage at node Y can be around the threshold voltage of $\text{M}_\text{X1}$ - $\text{M}_\text{X4}$. Then the leakage current through $\text{M}_\text{2}$ may become large enough to drive the $\text{M}_\text{6}$ and $\text{M}_\text{7}$ gate voltage low to bring up the voltage at node X and Y. Then the circuit starts to function as designed. For low temperature applications, the start-up may take a very long time. Therefore, we have added the start-up circuit on the right hand side.

Notice that at low current the Y node has a higher impedance than the X node, as the Y branch would not conduct current until the node voltage at Y reaches around the threshold voltage of $\text{M}_\text{X1}$ - $\text{M}_\text{X4}$. The X node sees a constant impedance of $\text{R}_\text{1}$. Then at high current, the Y node must see a lower impedance than the X node to maintain stability at DC. As described in~\cite{BGR_ZTC}, the transistors $\text{M}_\text{X1}$ - $\text{M}_\text{X4}$ operate at the edge of linear and saturation region to achieve zero temperature co-efficient (ZTC) current as well as ZTC drain voltage.

We consider that transistors $\text{M}_\text{X1}$ - $\text{M}_\text{X4}$ are effectively one long transistor $\text{M}_\text{X}$. We first assume $\text{M}_\text{X}$ works in saturation region and due to the amplifier formed by $\text{M}_\text{1}$ - $\text{M}_\text{5}$, the node voltage at X and Y will be the same. Thus, we reach the following expressions:

\begin{eqnarray}
\frac{1}{g_\text{m,sat}}&=&\frac{1}{\sqrt{2I_\text{D}\mu C_\text{ox}\frac{W}{L}}}\nonumber\\
                    R_\text{1}&=&\frac{2}{\sqrt{2I_\text{D}\mu C_\text{ox}\frac{W}{L}}}+\frac{V_\text{th}}{I_\text{D}}\nonumber\\
										&=&\frac{2}{g_\text{m,sat}}+\frac{V_\text{th}}{I_\text{D}}\,,
\label{eq:impedance_sat}
\end{eqnarray}

\noindent
where $g_\text{m,sat}$ is the transconductance of $\text{M}_\text{X}$ in saturation region and $I_\text{D}$ is the current in both of the X and Y branches. Similarly, if $\text{M}_\text{X}$ works in the linear region, then we can have:

\begin{eqnarray}
\frac{1}{g_\text{m,lin}}&=&\frac{1}{\mu C_\text{ox}\frac{W}{L}V_\text{DS}}\nonumber\\
                    R_\text{1}&=&\frac{1}{\mu C_\text{ox}\frac{W}{L}V_\text{DS}}+\frac{V_\text{DS}+2V_\text{th}}{2I_\text{D}}\nonumber\\
										&=&\frac{1}{g_\text{m,lin}}+\frac{V_\text{DS}+2V_\text{th}}{2I_\text{D}}\,,
\label{eq:impedance_lin}
\end{eqnarray}

\noindent
where $g_\text{m,lin}$ is is the transconductance of $\text{M}_\text{X}$ in linear region and $V_\text{DS}$ is the drain to source voltage of $\text{M}_\text{X}$. It can be seen that for both operating conditions, the impedance of Y branch is always smaller than that of X branch. Therefore, the Y and X branches form the positive and negative feedback loops respectively. For stable operations, we would like to ensure the negative feedback loop is always ``stronger'' than the positive feedback loop. That is to say, at all frequencies, the impedance at node Y should be smaller that the impedance seen at node X, which has already been indicated by (\ref{eq:impedance_sat}) and (\ref{eq:impedance_lin}) for DC. Frequency compensation of the two coupled loops can be performed similar to what is commonly done in low drop-out (LDO) voltage regulators, by adding zeros in the form of additional equivalent series resistance (ESR). The zeros from $\text{C}_\text{C1}$, $\text{R}_\text{C1}$ and $\text{C}_\text{C2}$, $\text{R}_\text{C2}$ cancels the internal pole at the gate of $\text{M}_\text{6}$ and $\text{M}_\text{7}$ for individual loops. The compensation load capacitors $\text{C}_\text{C1}$ and the impedance seen at node X create an output pole $p_\text{O1}$ at this node. Similarly another output pole $p_\text{O2}$ can be identified at node Y. It is obvious that $p_\text{O2}$ frequency is lower than that of $p_\text{O1}$, which maintains stability of the entire circuit.

We simulated this circuit and obtained the following performance and sensitivity parameters as shown in Table \ref{tb:BGR_ZTC_performance} and Table \ref{tb:BGR_ZTC_sensitivity} respectively. It can be seen from Table \ref{tb:BGR_ZTC_sensitivity} that the output reference voltage is closely related to process parameters such as the threshold voltage of $\text{M}_\text{X}$ and the absolute value of the on-chip resistors. These two parameters change at low temperatures. In addition, the matching for resistor $\text{R}_\text{2a}$ is also important to the designs. For those reasons, we implemented binary based trimmings to both $\text{R}_\text{1}$ and $\text{R}_\text{2a}$ in Fig. \ref{fig:BGR_ZTC} and the LSB accuracy is about 1\% of the individual resistor value. The back-gate control node can also be used to tune the threshold voltage of $\text{M}_\text{X}$ at low temperature. We note that the simulated nominal output reference voltage is now \unit[494]{mV} after implementing trimmings.

\begin{figure}
\centering
\vspace*{-0cm}
\hspace*{-0.2cm}
\includegraphics[width=0.45\textwidth, angle=0]{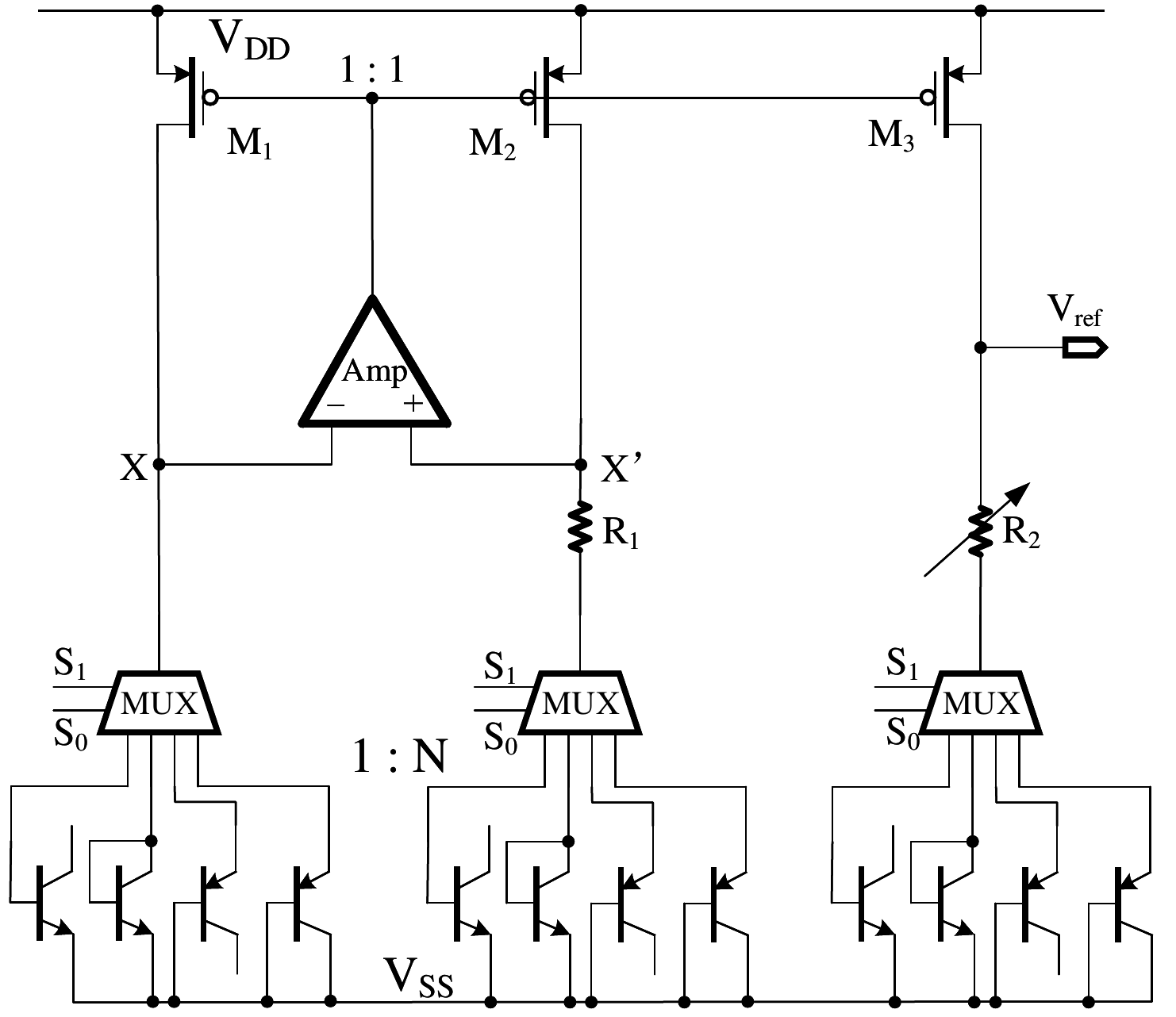}
\vspace*{-0.2cm}
\caption{BJT based BGR in \unit[28]{nm} FDSOI CMOS process. Four terminal multiplexers are used to connect different types of BJTs or PN junctions into the circuit.}
\vspace*{-0cm}
\label{fig:BGR_BJT}
\end{figure}

In this \unit[28]{nm} FDSOI CMOS process, we have also implemented a conventional BJT based BGR circuit as a experimental comparison. Due to the increase in threshold voltage in MOS transistors, we shifted the cascoded current mirror BGRs into a low supply voltage oriented design, as shown in Fig. \ref{fig:BGR_BJT}. This circuit is powered by \unit[1.8]{V} supply. Multiplexers are used to switch among different types of BJTs (the lateral PNP or vertical NPN) and PN junctions.

\begin{table}
\setlength{\extrarowheight}{2pt}
\caption{Performances summary of MOS-only voltage reference circuit}
\label{tb:BGR_ZTC_performance}
\vspace*{0cm}
\centering
\begin{tabular}{|c|c|c|}
\hline
\multicolumn{2}{|c|}{Performance} & Simulation results \\
\hline
\multicolumn{2}{|c|}{Nominal $V_\text{REF}$} & \unit[503]{mV} \\
\hline
\multicolumn{2}{|c|}{Minimum $V_\text{DD}$} & \unit[0.8]{V} \\
\hline
\multicolumn{2}{|c|}{Temperature range} & \unit[-40]{$^{\circ}$C} - \unit[150]{$^{\circ}$C} \\
\hline
\multicolumn{2}{|c|}{Temperature co-efficiency} & \unit[1.3]{ppm/$^\circ$C} \\
\hline
\multirow{5}{*}{Quiescent current} & \unit[150]{$^{\circ}$C} & \unit[9.1]{$\mu$A} \\
                                   & \unit[25]{$^{\circ}$C} & \unit[8.8]{$\mu$A} \\
																	 & \unit[-40]{$^{\circ}$C} & \unit[8.6]{$\mu$A} \\
																	 & \unit[-196]{$^{\circ}$C} (\unit[77]{K}) & \unit[7.7]{$\mu$A} \\
																	 & \unit[-269]{$^{\circ}$C} (\unit[4]{K}) & \unit[7.7]{$\mu$A} \\	
\hline
\multirow{5}{*}{PSR @ \unit[25]{$^{\circ}$C}} & DC & \unit[136.6]{dB} \\
                     & \unit[10]{kHz} & \unit[136.2]{dB} \\  
										 & \unit[100]{kHz} & \unit[124.4]{dB} \\
										 & \unit[1]{MHz} & \unit[82.2]{dB} \\
										 & \unit[10]{MHz} & \unit[37.5]{dB} \\
\hline
\multirow{5}{*}{PSR @ \unit[4]{K}} & DC & \unit[130.0]{dB} \\
                     & \unit[10]{kHz} & \unit[117.0]{dB} \\  
										 & \unit[100]{kHz} & \unit[96.3]{dB} \\
										 & \unit[1]{MHz} & \unit[58.3]{dB} \\
										 & \unit[10]{MHz} & \unit[22.1]{dB} \\
\hline
\end{tabular}
\vspace*{-0cm}
\end{table}

\begin{table}
\setlength{\extrarowheight}{2pt}
\caption{Process sensitivities of MOS-only voltage reference circuit}
\label{tb:BGR_ZTC_sensitivity}
\vspace*{0cm}
\centering
\begin{tabular}{|c|c|c|}
\hline
\multicolumn{2}{|c|}{Error sources} & $V_\text{REF}$ deviations \\
\hline
$V_\text{th}$ vary by \unit[1]{mV} & $\text{M}_\text{X}$ & \unit[1]{mV} \\
\hline
\multirow{3}{*}{$V_\text{th}$ mismatch by \unit[1]{mV}} & $\text{M}_\text{1}$ and $\text{M}_\text{2}$ & \unit[0.15]{mV} \\
																											  & $\text{M}_\text{3}$ and $\text{M}_\text{4}$ & \unit[0.11]{mV} \\
																											  & $\text{M}_\text{6}$ and $\text{M}_\text{7}$ & \unit[1]{mV} \\
\hline
Resistor values vary by 1\% & $\text{R}_\text{1}$, $\text{R}_\text{2a}$ and $\text{R}_\text{2b}$ & \unit[1.1]{mV} \\
\hline
\multirow{3}{*}{Resistor mismatch by 1\%} & $\text{R}_\text{1}$ & \unit[0.45]{mV} \\
																					& $\text{R}_\text{2a}$ & \unit[1.6]{mV} \\
																					& $\text{R}_\text{2b}$ & Negligible \\
\hline																					
\end{tabular}
\vspace*{-0cm}
\end{table}

\begin{figure*}[!t]
\centering
\vspace*{-0.2cm}
\hspace*{-0cm}
\includegraphics[width=0.8\textwidth, angle=0]{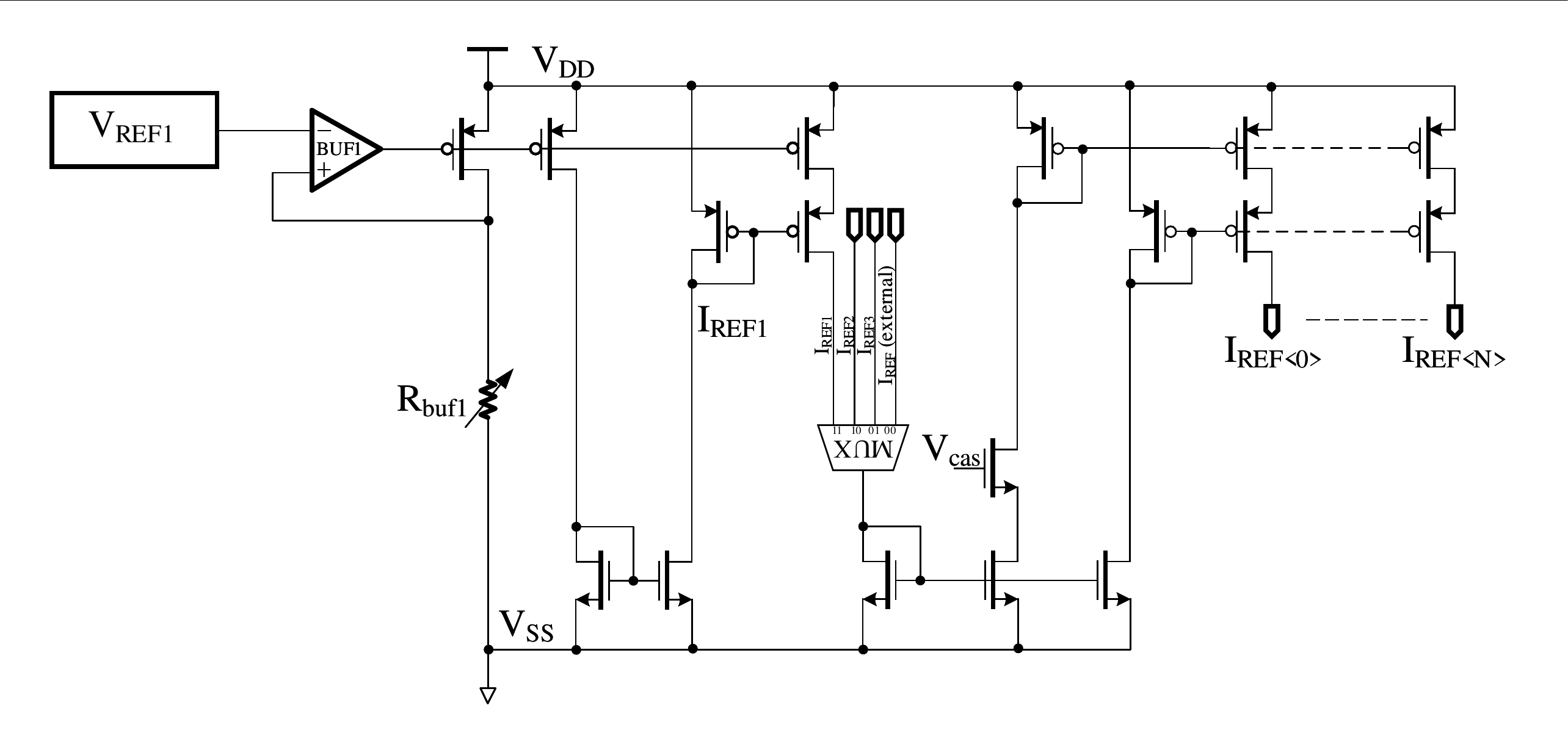}
\vspace*{-0.2cm}
\caption{Simplified current reference circuit in \unit[28]{nm} FDSOI CMOS process. The reference is generated via buffering a voltage on a resistor. A multiplexing scheme is adopted to have multiple choice of the reference current. The output current can be routed for other blocks as well as for testing.}
\vspace*{-0cm}
\label{fig:IREF}
\end{figure*}

The current reference provides bias current for most of the analog sub-modules of the chip. The current reference is designed using existing voltage reference topologies. We have implemented a conventional BJT based BGR, a BJT based BGR current generator and a MOS-only voltage reference for this current reference. The simplified diagram is shown in Fig. \ref{fig:IREF}, which only shows one version of voltage to current conversion. In this design, we employ a multiplexing scheme as we expect the BGR circuits using normal silicon BJTs will freeze-out at temperatures below \unit[60]{K}. Due to the negative TC of the on-chip resistors, the voltage references cannot be directly plugged in for voltage to current conversion. In fact, we have generated a CTAT voltage and applied it to the resistor $\text{R}_\text{buf1}$ to create a ZTC reference current. Resistor trimming has been used, for example,  $\text{R}_\text{2}$ in BGR circuit and $\text{R}_\text{1}$ in MOS-only reference circuit. The trimming range of $\text{R}_\text{buf1}$ covers approximate $\pm$60\% of the range with Least Significant Bit (LSB) accuracy of 2\%.

\section{Cryogenic Measurements}

\begin{figure}
\centering
\vspace*{-0cm}
\hspace*{-0.2cm}
\includegraphics[width=0.45\textwidth, angle=0]{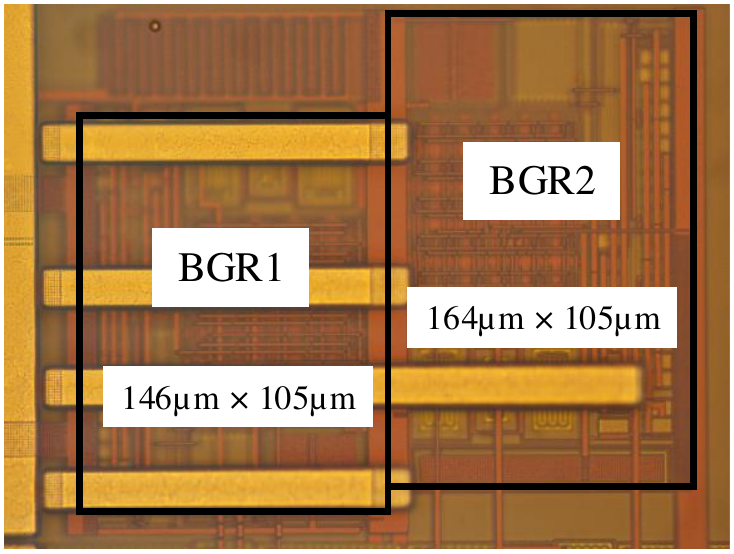}
\vspace*{-0.2cm}
\caption{BGR circuits implemented in \unit[0.35]{$\mu$m} SiGe BiCMOS process.}
\vspace*{-0cm}
\label{fig:SiGe_BGRs}
\end{figure}

\begin{figure}
\centering
\vspace*{-0cm}
\hspace*{-0.2cm}
\includegraphics[width=0.45\textwidth, angle=0]{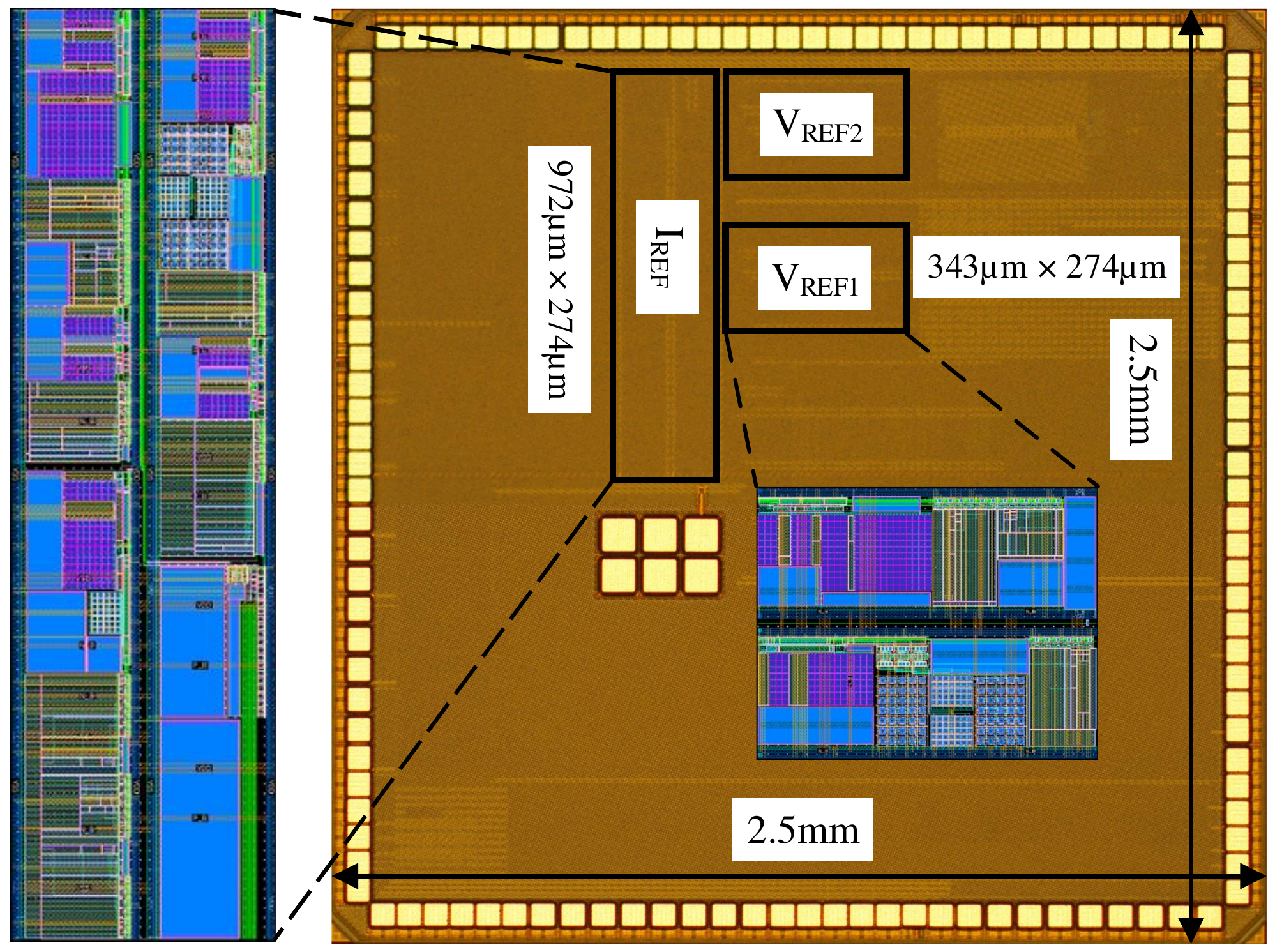}
\vspace*{-0.2cm}
\caption{Voltage and current reference circuits fabricated in \unit[28]{nm} FDSOI CMOS process. We have implemented both long channel and short channel voltage reference circuits and the current reference circuit is implemented with only short channel devices.}
\vspace*{-0cm}
\label{fig:GB}
\end{figure}

We fabricated two BGR-based voltage reference circuits in \unit[0.35]{$\mu$m} SiGe BiCMOS process. The chip photo is shown in Fig. \ref{fig:SiGe_BGRs}. BGR1 is a conventional SiGe HBT based current mirror BGR and the BGR2 has an output buffer stage, (see Fig. \ref{fig:BGRs_SiGe} for schematic diagrams). We have also implemented four voltage reference circuits and three current references in a \unit[28]{nm} FDSOI CMOS process.  The chip photo is presented in Fig. \ref{fig:GB}, in which, we have also displayed the layout of the circuits as the chip is fully covered by density fills. In this figure, $\text{V}_\text{REF1}$ is the long channel voltage references and $\text{V}_\text{REF2}$ is the short channel version.

The SiGe BGRs are tested in a LakeShore CRX-4K closed cycle refrigerator (CCR) based cryogenic probe station with LakeShore TC336 temperature monitor and control unit. This setup enables a wide temperature range sweep from \unit[250]{K} to \unit[6]{K}. The probe station is not fully enclosed by the radiation shield limiting the base temperature to \unit[6]{K}. The advantage of using a cryogenic probe station is fast turn-around time (cools down and warms up quickly due to small thermal mass) and temperature control for the test samples. The temperature of the probe station can be set and stablized to a given value.

\begin{figure}[!b]
\centering
\vspace*{-0cm}
\hspace*{-0.2cm}
\includegraphics[width=0.5\textwidth, angle=0]{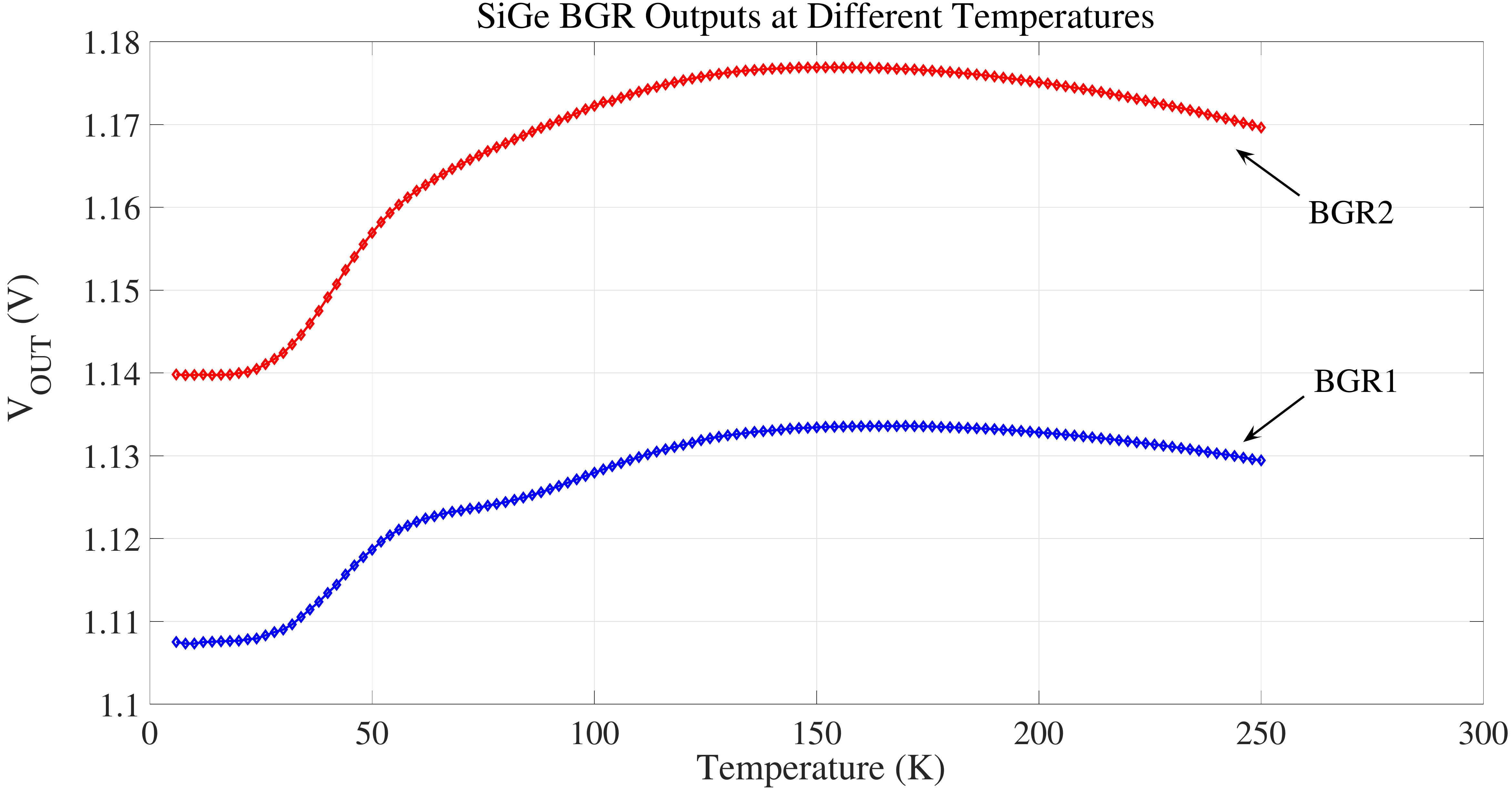}
\vspace*{-0.2cm}
\caption{Measured SiGe BGRs output voltage along temperature sweep from \unit[250]{K} to \unit[6]{K}.}
\vspace*{-0cm}
\label{fig:SiGe_BGRs_tempsweep}
\end{figure}

\begin{figure*}[!t]
\centering
\vspace*{-0.2cm}
\hspace*{-0cm}
\includegraphics[width=1\textwidth, angle=0]{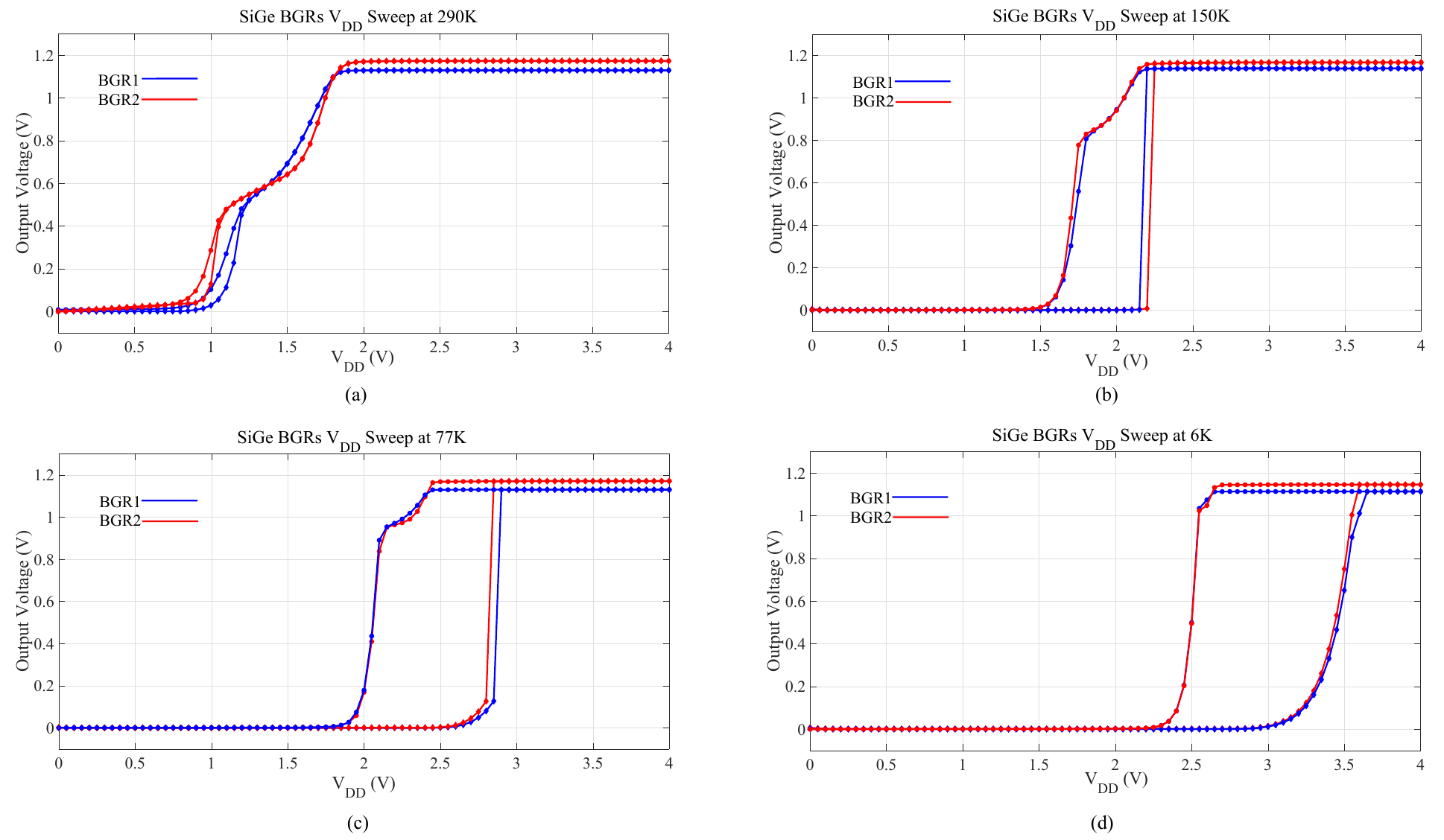}
\vspace*{-0.2cm}
\caption{Power supply sweep tests on BGR1 and BGR2 circuits at (a), \unit[290]{K}; (b), \unit[150]{K}; (c), \unit[77]{K} and (d), \unit[6]{K}. Larger hysteresis is observed at lower temperatures.}
\vspace*{-0cm}
\label{fig:BGR_VddSweep}
\end{figure*}

We have measured the SiGe BGRs and plotted the output voltage with respect to temperature in Fig \ref{fig:SiGe_BGRs_tempsweep}. The output voltage is recorded using Keysight 34460A digital multi meter (DMM) with load impedance set to \unit[1]{G$\Omega$} to minimize loading to the BGR circuits. The room-temperature current consumptions of BGR1 and BGR2 are \unit[4]{$\mu$A} and \unit[4.8]{$\mu$A} respectively from a \unit[3.3]{V} power supply. At low temperature (\unit[6]{K}), we need to increase the power supply voltage to \unit[3.6]{V} due to increased threshold voltages in both PMOS and NMOS transistors. The currents consumed by BGR1 and BGR2 reduce to about \unit[0.4]{$\mu$A} and \unit[0.5]{$\mu$A} respectively, corresponding to power consumption of approximate \unit[1.5]{$\mu$W} and \unit[1.8]{$\mu$W} respectively. We believe this is due to the reduction in PTAT currents in BGRs. The maximum deviation of output voltage from BGR1 is \unit[26]{mV} across the measured temperature range (\unit[6]{K} to \unit[250]{K}), resulting in a TC of \unit[107]{ppm/$^\circ$C}. BGR2 has a maximum reference output voltage change of \unit[37]{mV}, indicating a TC of \unit[152]{ppm/$^\circ$C}. These temperature stability performances are comparable with the one in~\cite{sub-1K-BGR-2009} and our BGRs consumes much less power at cryogenic temperature (\unit[1.5]{$\mu$W} and \unit[1.8]{$\mu$W} versus \unit[130]{$\mu$W}).

Power-supply sweep tests are performed at 4 different temperatures as shown in Fig. \ref{fig:BGR_VddSweep}. We observe that the BGR circuits start to work at higher supply voltages when the temperature is lowered, as the threshold voltage increases at low temperatures. The hysteresis between power-up and power-down becomes large at low temperatures. At \unit[6]{K}, low power-supply design techniques may be necessary for many of the analog designs.

\begin{figure}
\centering
\vspace*{-0.2cm}
\hspace*{-0.1cm}
\includegraphics[width=0.5\textwidth, angle=0]{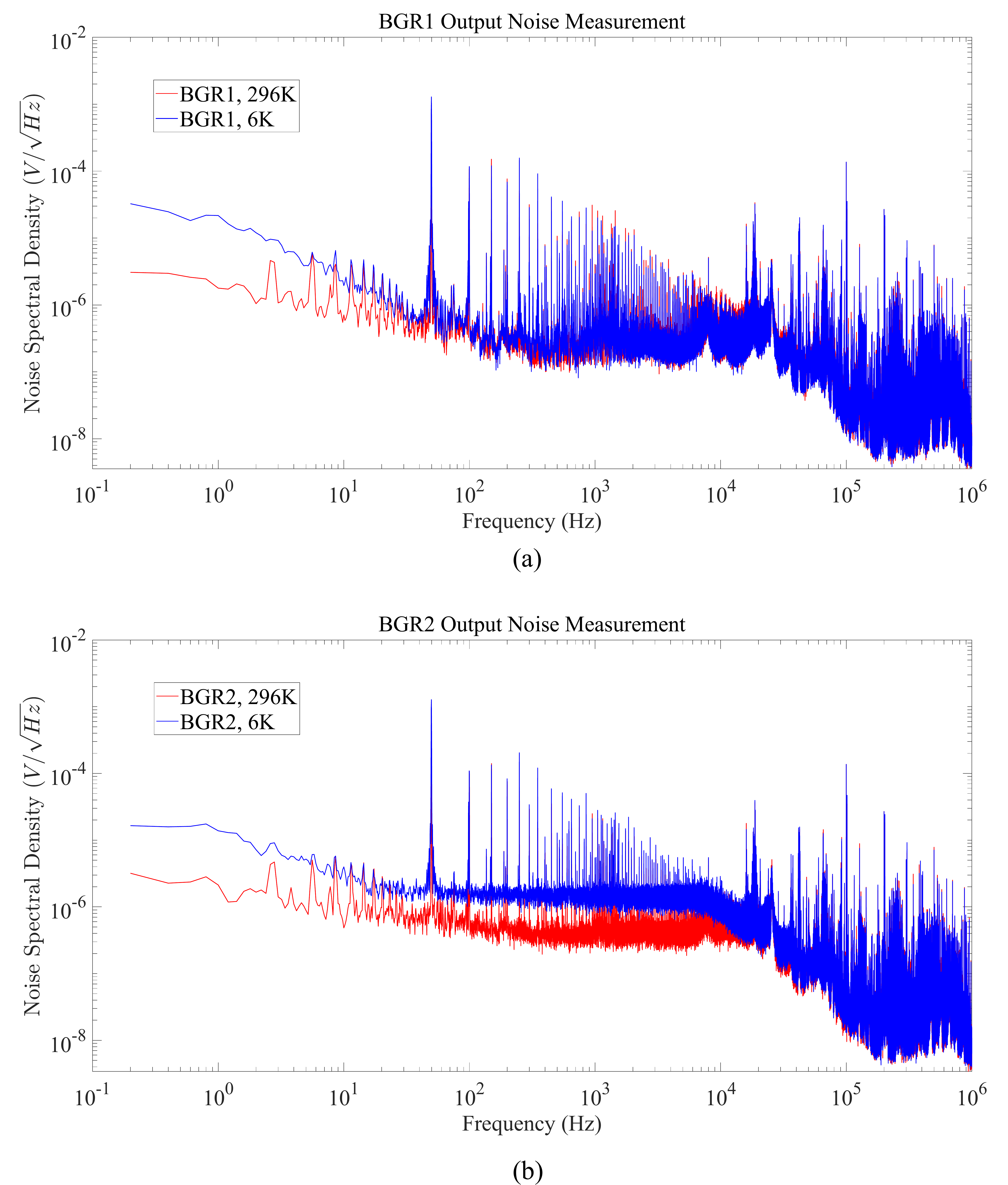}
\vspace*{-0.2cm}
\caption{Measured SiGe BGRs' noise performance at room temperature and $\unit[6]{K}$. (a), BGR1 noise performance; (b), BGR2 noise performance.}
\vspace*{-0cm}
\label{fig:BGRs_Noise}
\end{figure}

The noise performance of the SiGe BGR circuits are shown in Fig. \ref{fig:BGRs_Noise}. In this experiment, the output of the BGRs are connected to Stanford research systems (SR560). The output of the amplifier is captured using a digital scope. The purpose of using this setup is that the BGRs cannot directly drive the input load of the scope. We can see that at low temperature, the low-frequency noise of both of the two BGRs are higher than their room-temperature counterparts. BGR1 and BGR2 show similar room-temperature noise performance, and at cryogenic temperatures, the 1/f noise from the MOS transistors, especially from the current mirrors increases the output noise. At low frequencies, BGR1 and BGR2 have similar noise characteristics. The BGR2 shows a higher noise power spectral density at the mid-frequency range at low temperature compared to BGR1. It is likely that at low temperature, due to drastically reduced bias current in all circuit branches, the loop bandwidth of the output buffer in BGR2 becomes too narrow to provide efficient feedback. Although a compensation capacitor is implemented in BGR2 output buffer stage, see Fig. \ref{fig:BGRs_SiGe}(b), it is not effective to provide a noise coupling path at this mid-frequency range. Therefore, the noise current from $\text{M}_\text{7}$ and $\text{M}_\text{8}$ cascode sees different impedance at the drain side of $\text{M}_\text{8}$. In BGR1, this impedance is approximately $\text{R}_\text{2}$ plus the impedance of diode connected BJT. In BGR2, the noise current cannot pass through the base of $\text{Q}_\text{3}$, because when there is a small base current, $\text{Q}_\text{3}$ will source a large current from the supply to bring up the emitter voltage. Therefore, this noise current sees roughly a collector-emitter resistance, which is typically higher than that in BGR1, and converts into a noise voltage. This noise voltage is then passed to the output via emitter-follower configured $\text{Q}_\text{3}$. At higher frequencies, the compensation coupling capacitor begins to be effective and a noise cancellation path via $\text{C}_\text{C}$ and $\text{Q}_\text{4}$ can be identified.

\begin{figure}
\centering
\vspace*{-0.2cm}
\hspace*{-0.2cm}
\includegraphics[width=0.5\textwidth, angle=0]{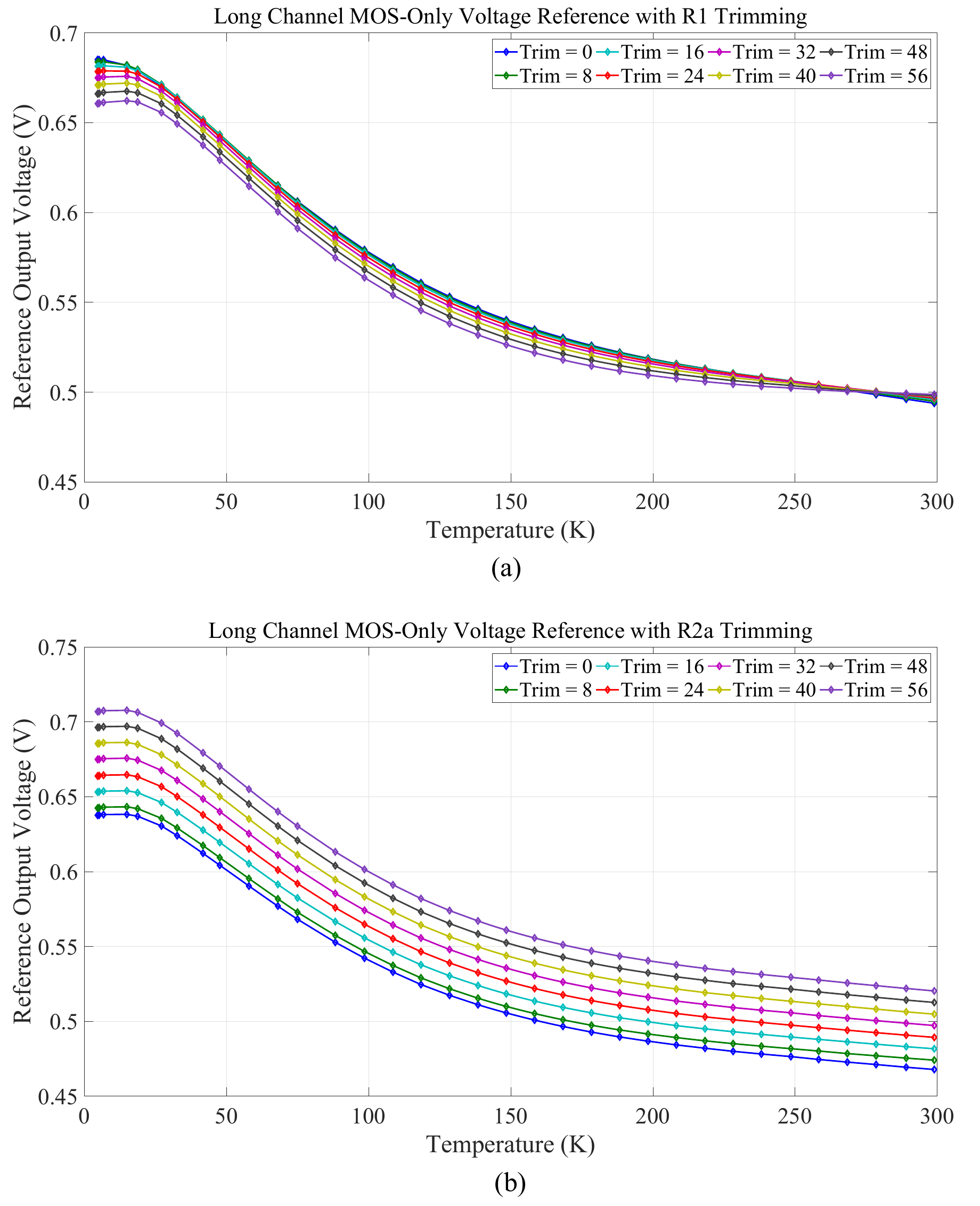}
\vspace*{-0.2cm}
\caption{Measurement results on resistor trimmings in the long channel MOS-only voltage reference circuit with zero back-gate control voltage: (a), $\text{R}_\text{1}$ trimming with $\text{R}_\text{2a}$ set to its nominal value, the bigger the trimming code number, the lower the resistance; (b), $\text{R}_\text{2a}$ trimming with $\text{R}_\text{1}$ set to its nominal value, the bigger the trimming code number, the lower the resistance.}
\vspace*{-0cm}
\label{fig:LC_MOS_Trim}
\end{figure}

The reference circuits taped-out in \unit[28]{nm} FDSOI CMOS technology are measured in a BlueFors XLD dilution refrigerator. In this setup, we first acquire data at room temperature, then we cool down the sample in the fridge and perform measurements whenever a predetermined temperature is reached. For the long channel MOS-only voltage reference circuit trimming test, we trim $\text{R}_\text{1}$ and $\text{R}_\text{2a}$ and apply zero back-gate control voltage, (see Fig. \ref{fig:BGR_ZTC} for circuit diagram information). The measurement results are shown in Fig. \ref{fig:LC_MOS_Trim}. We implement 6-bit trimming for both of these resistors. Due to the large threshold voltage increase at low temperatures, the predetermined trimming range is not sufficient to bring the reference voltage back to its room-temperature value.

\begin{figure}[!b]
\centering
\vspace*{-0.2cm}
\hspace*{-0.2cm}
\includegraphics[width=0.5\textwidth, angle=0]{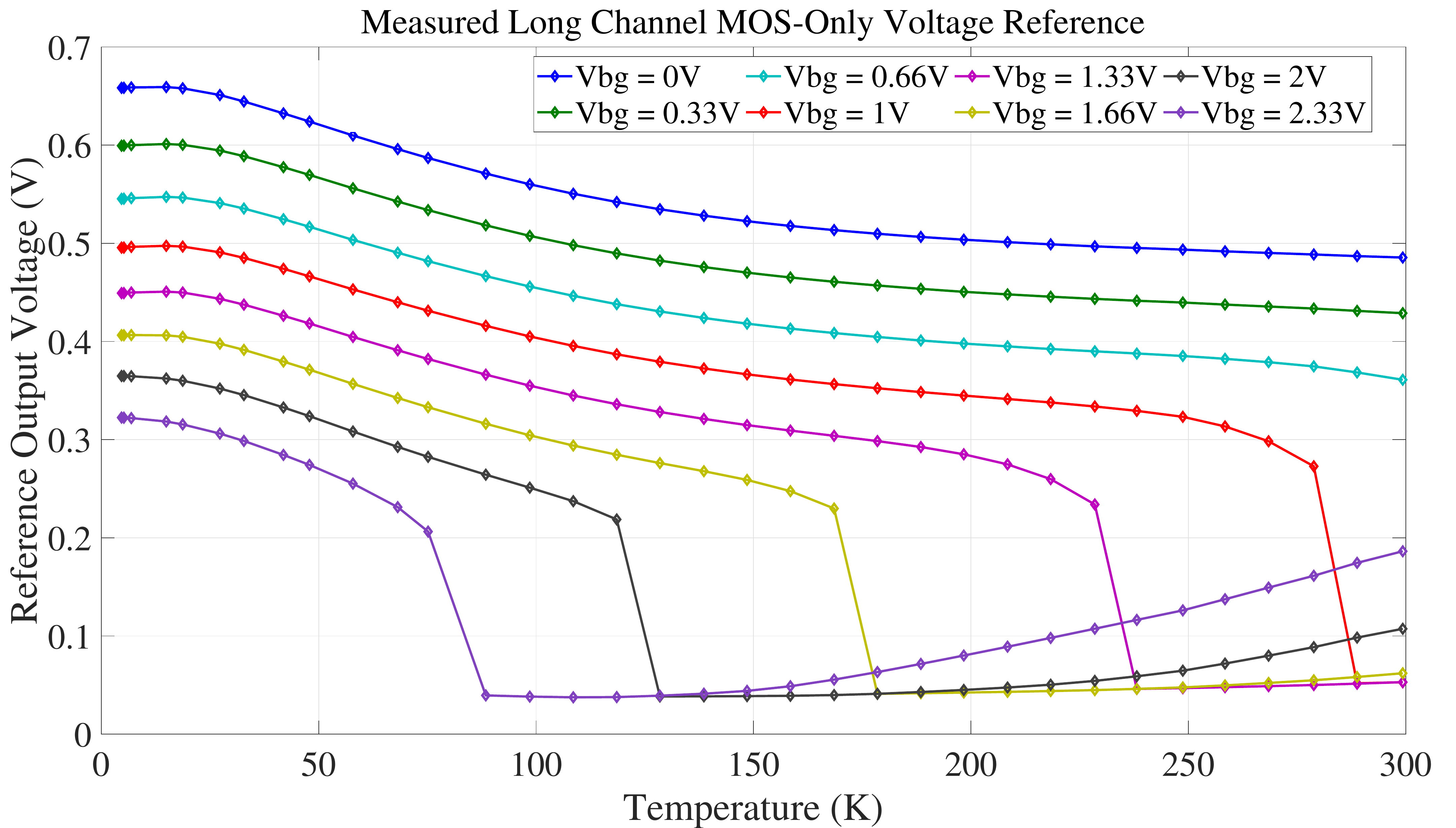}
\vspace*{-0.2cm}
\caption{Measurement results on long channel MOS-only voltage reference circuit for different back-gate control voltages along with the temperature sweep. The back-gate control voltage is noted as ``Vbg'' in the figure.}
\vspace*{-0cm}
\label{fig:LC_MOS}
\end{figure}

\vfill

The back-gate control has a stronger effect on the output voltage of this reference circuit as it directly modulates the threshold voltage of $\text{M}_\text{N1}$ - $\text{M}_\text{N4}$ transistors. The default or zero back-gate bias in this process is the ground potential for both of the PMOS and NMOS transistors. When we refer to back-gate control voltage in our measurement, unless particularly pointed out, we use a positive bias voltage for the NMOS and a negative (with respect to ground) voltage for the PMOS transistors. The measured long channel voltage reference for different back-gate control voltages is presented in Fig. \ref{fig:LC_MOS}. It can be seen that at \unit[4]{K}, applying \unit[1]{V} back-gate voltage can bring the output voltage of this circuit to its room temperature value. We can also observe from Fig. \ref{fig:LC_MOS} that the output voltage in \unit[4]{K} - \unit[20]{K} temperature range shows a relative low TC (the voltage curve is ``flatter''). This is important for our application as the chip can suffer from self-heating effects resulting from higher thermal resistance between the chip and cryostat at these temperatures.

\begin{figure}
\centering
\vspace*{-0.2cm}
\hspace*{-0.2cm}
\includegraphics[width=0.25\textwidth, angle=0]{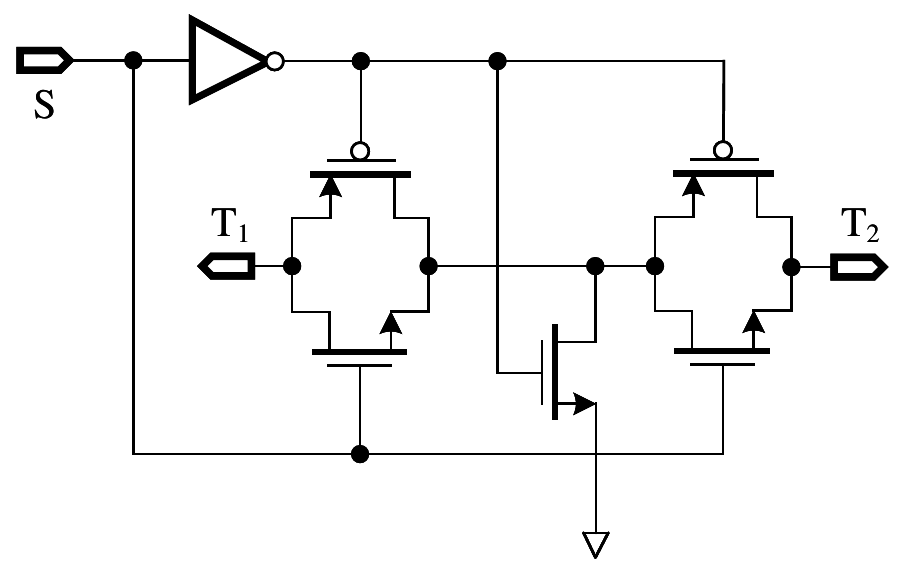}
\vspace*{-0.2cm}
\caption{Transistor level switch diagram used in ATM MUX. The ground-connected NMOS transistor shunts off-switch noise when two transmission gates are turned OFF.}
\vspace*{-0cm}
\label{fig:ATM_SW}
\end{figure}

With the testing architecture implemented via ATM(s), T-switches are used for better noise performance. A simplified switch schematic is displayed in Fig. \ref{fig:ATM_SW}. At room temperature, the NMOS transistor cannot be fully turned OFF when a back-gate voltage of more than \unit[1]{V} is applied, as this shorts the output to the ground. Thus we see a zero output voltage. When the temperature is lowered, this switch can be turned OFF with higher back-gate voltage. For back-gate voltage greater than \unit[2.66]{V}, even at cryogenic temperatures, this NMOS transistor fails to turn OFF. Also it is interesting to notice for high back-gate biasing at room temperatures, the output voltage is not zero. This is because the transistors stay ON and  the cascode PMOS current mirror shorts the power supply to the shunt NMOS transistors. This forms a resistive voltage divider and the output voltage is the divided power supply voltage.

\begin{figure}
\centering
\vspace*{-0.2cm}
\hspace*{-0.2cm}
\includegraphics[width=0.5\textwidth, angle=0]{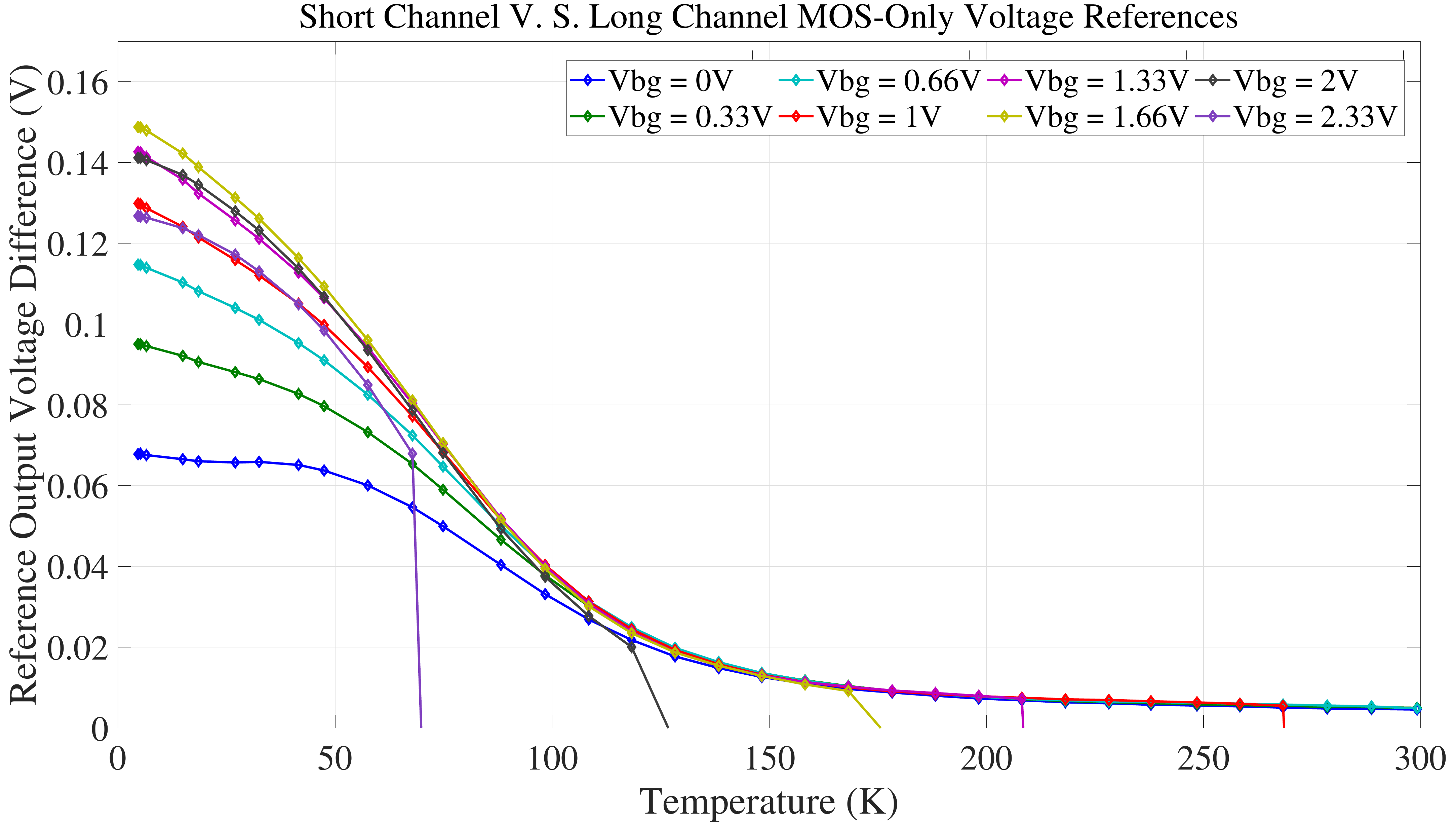}
\vspace*{-0.2cm}
\caption{Short channel version MOS-only reference voltage minus that of long channel version.}
\vspace*{-0cm}
\label{fig:SCVSLC_MOS}
\end{figure}

The short channel version of the MOS-only voltage reference exhibits a similar behaviour. The difference mainly lies in the low-temperature part of the measurement. Figure \ref{fig:SCVSLC_MOS} shows the difference between these two circuits. The short channel MOS-only reference circuit generally has a higher output voltage than the long channel version. The output voltage difference reaches a peak value at \unit[1.66]{V} back-gate voltage and declines for higher back-gate biases. This result can be experimentally reproduced across different sample chips, likely due to the excessive number (32 in our case) of short-channel transistors implemented for $\text{M}_\text{X1}$ - $\text{M}_\text{X4}$ . At low-temperatures, the source and drain resistance may potentially increase. We can see at higher back-gate voltages, while the threshold voltage of the output transistor decreases, the increase of source and drain resistance takes a larger portion of the output voltage, which results in larger voltage differences between the short-channel and long-channel voltage references. Further increase back-gate voltages, i. e. from \unit[1.66]{V} and above, referring to Fig. \ref{fig:BGR_ZTC}, the output transistors' threshold voltage will reduce and thus the current in both X and Y branches. The amount of voltage drop on the increased source and drain resistance tends to decrease which results in output voltage different decrease.

\begin{figure}[!b]
\centering
\vspace*{-0.2cm}
\hspace*{-0.2cm}
\includegraphics[width=0.5\textwidth, angle=0]{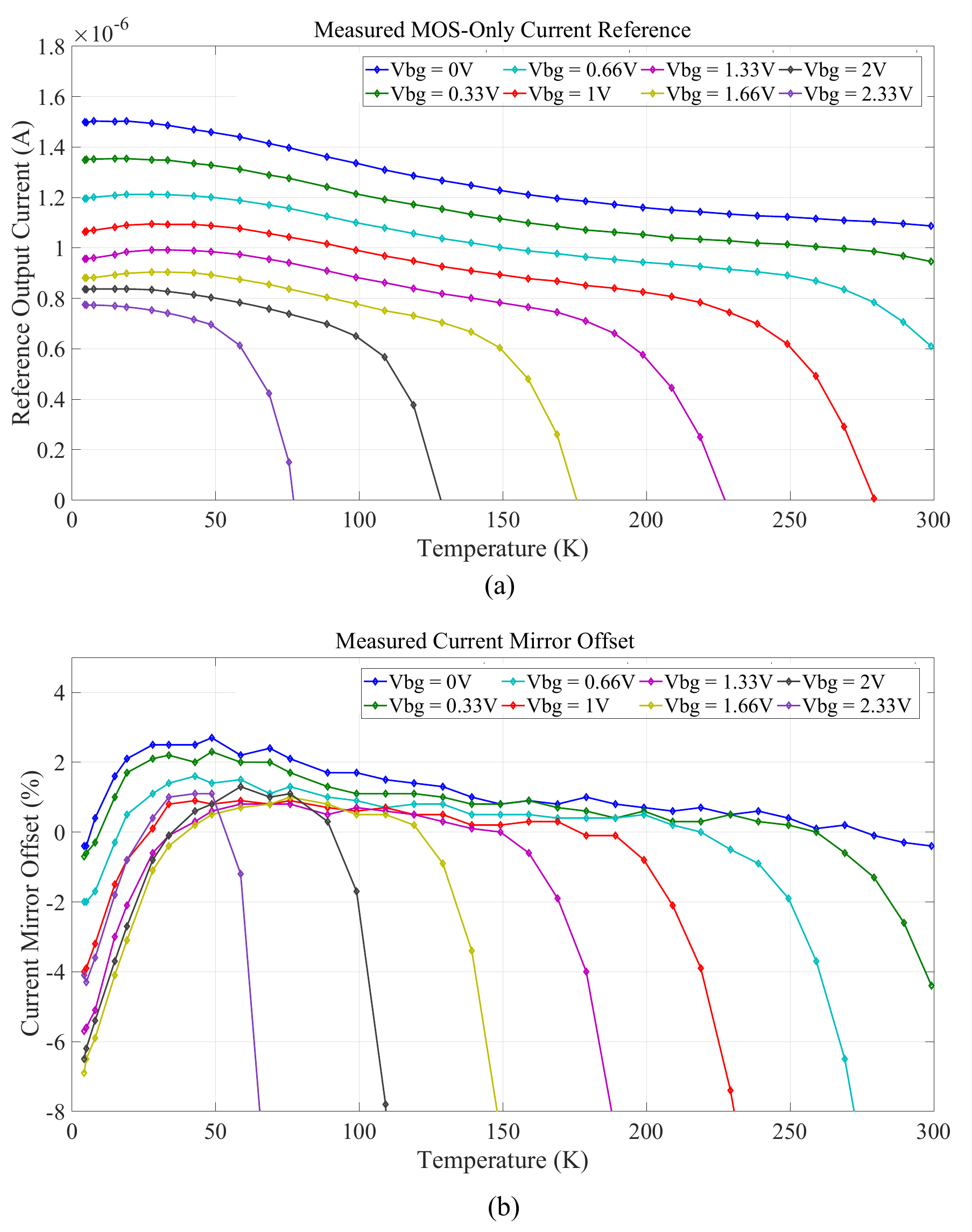}
\vspace*{-0.2cm}
\caption{Measured current reference at different temperatures using back-gate controls. (a), reference output using nominal trimming values for $\text{R}_\text{1}$ and $\text{R}_\text{buf1}$; (b), measured current mirror offset at different back-gate biases. The offset is measured by applying an external \unit[1]{$\mu$A} input current and subtracting the input current from the output current. A negative value indicates the output current is smaller than the input current.}
\vspace*{-0cm}
\label{fig:CG_CM_Mismatch}
\end{figure}

\begin{figure}
\centering
\vspace*{-0.2cm}
\hspace*{-0.2cm}
\includegraphics[width=0.5\textwidth, angle=0]{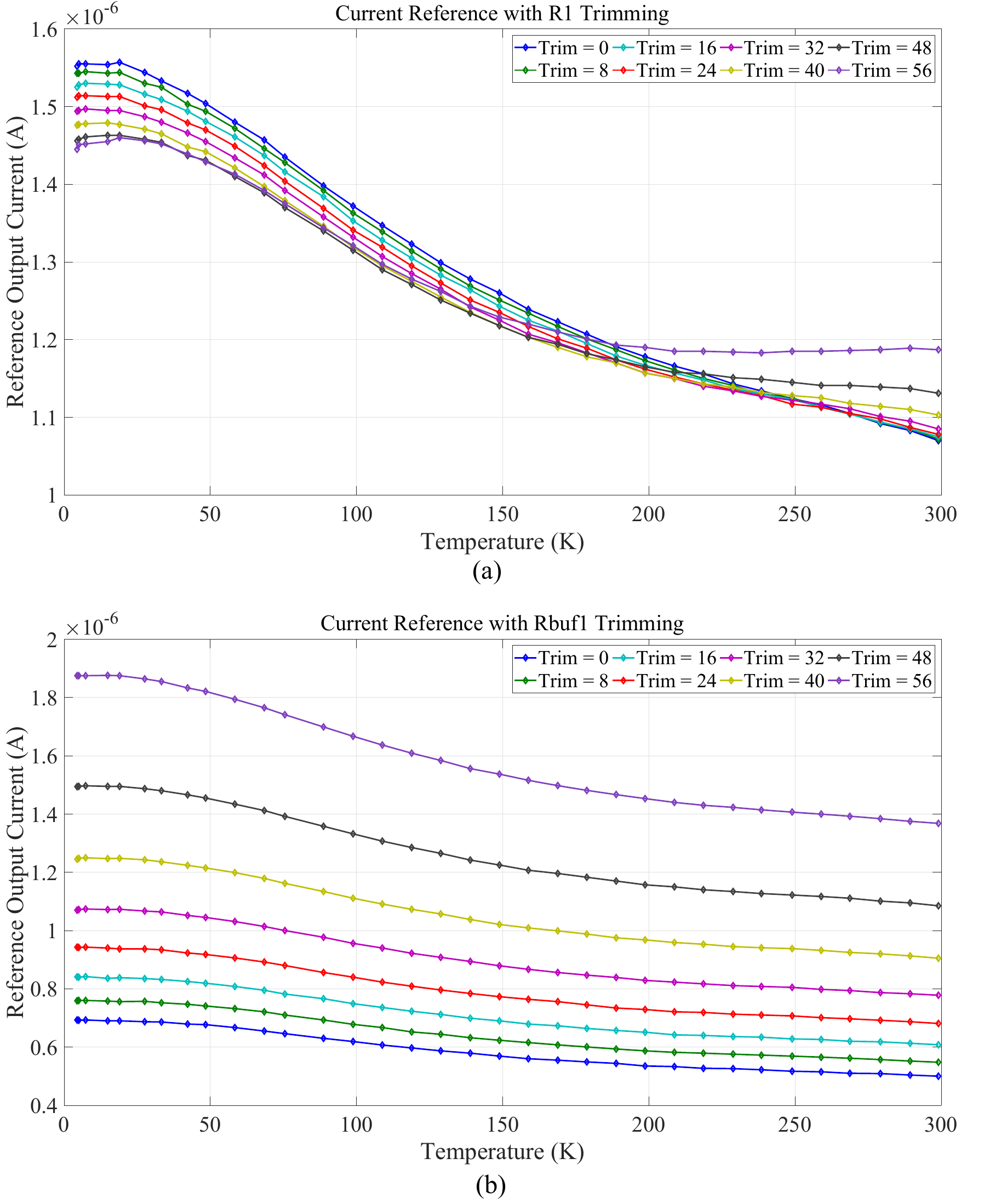}
\vspace*{-0.2cm}
\caption{Resistor trimming measurement in MOS-only current reference and a larger trimming code is corresponding to a lower resistance value. In this measurement, we apply zero back-gate bias. (a), $\text{R}_\text{1}$ trimming when $\text{R}_\text{buf1}$ is set to its nominal value; (b), $\text{R}_\text{buf1}$ trimming when $\text{R}_\text{1}$ is set to its nominal value. Both of $\text{R}_\text{1}$ and $\text{R}_\text{buf1}$ have 6-bit binary weighted trimming accuracy.}
\vspace*{-0cm}
\label{fig:CG_MOS_Trim}
\end{figure}

The MOS-only current reference shows similar trend as the voltage reference circuit, shown in Fig. \ref{fig:CG_CM_Mismatch}(a). The main reason is the similar circuit topology used in Fig. \ref{fig:IREF}. The current mirror, which is the last stage of the current reference, is also tested for its offset characteristics using an external \unit[1]{$\mu$A} current. The difference is taken via subtracting the input current from the output current. The large negative values (which are off the scale of this plot) are mainly due to the ATM MUX switches, which short the measurement instrument to the chip ground at high back-gate voltages. We can see that at low end of the measured temperature range, the output current tends to have a smaller values of a given input current, perhaps due to the matching property changing in the MOS transistors for a high back-gate bias. The offset reaches a peak of approximate 7.5\% at \unit[1.66]{V} back-gate control voltage. 

We have also performed resistor trimming test during the temperature sweep. In this test, we use MOS-only voltage reference in Fig. \ref{fig:BGR_ZTC} to drive the current generator circuit (Fig. \ref{fig:IREF}) and we trim $\text{R}_\text{1}$ and $\text{R}_\text{buf1}$ independently. Trimming $\text{R}_\text{2a}$ resistor (Fig. \ref{fig:BGR_ZTC}) has a similar effect as trimming $\text{R}_\text{buf1}$, as both of them shift the curves vertically, but $\text{R}_\text{buf1}$ has a much wider trimming range. The resistor trimming experiment results are presented in Fig. \ref{fig:CG_MOS_Trim}. In this measurement, we apply zero back-gate bias. 

\begin{figure}[!t]
\centering
\vspace*{-0.2cm}
\hspace*{-0.2cm}
\includegraphics[width=0.5\textwidth, angle=0]{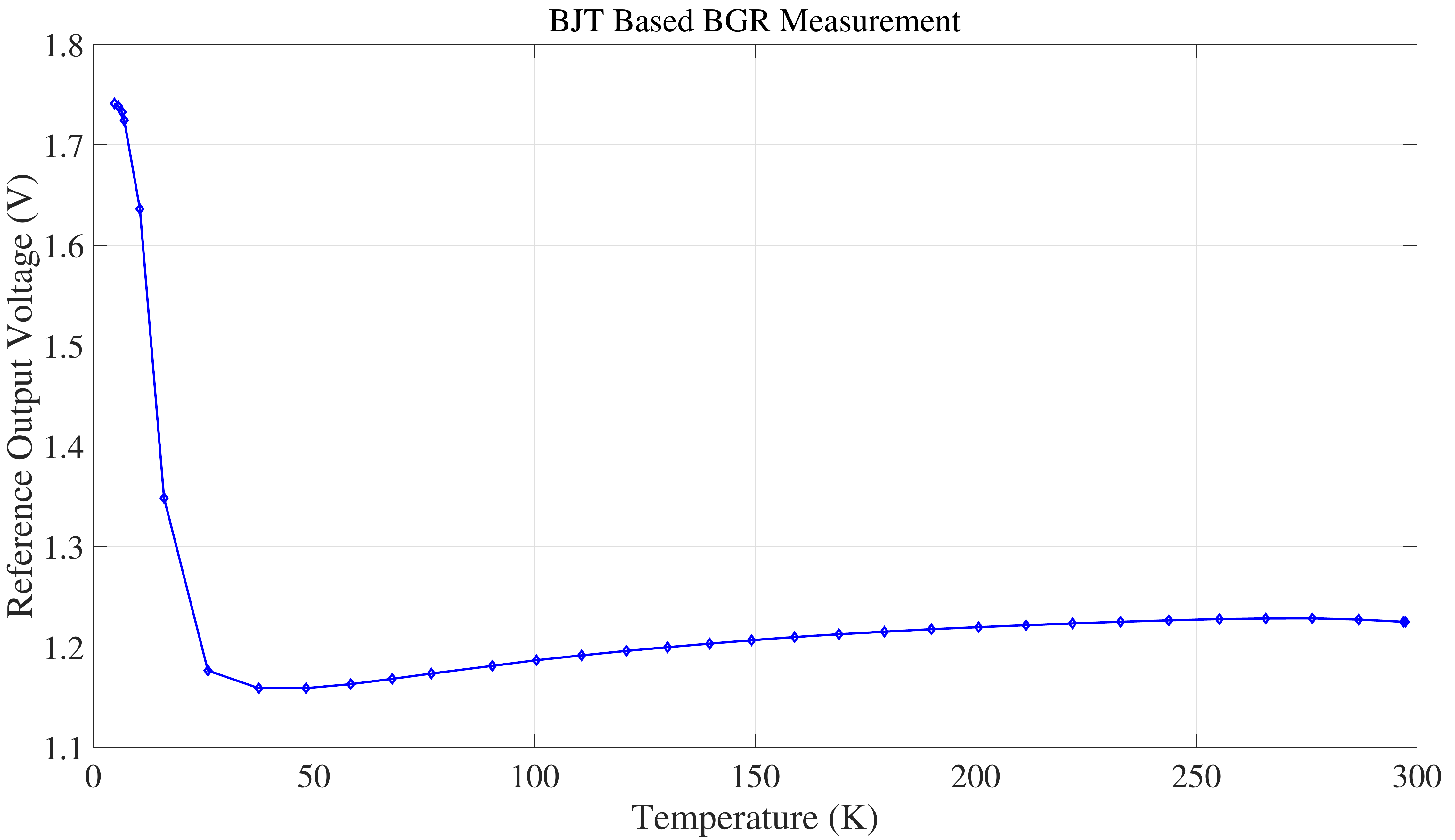}
\vspace*{-0.2cm}
\caption{Measurement of long channel BJT based BGR in \unit[28]{nm} FDSOI CMOS process, the PNP BJTs are used in this test. The short channel and NPN version of the circuit show similar results.}
\vspace*{-0cm}
\label{fig:BGR_BJT_FDSOI}
\end{figure}

The silicon BJT based BGRs in this \unit[28]{nm} FDSOI process fail at about \unit[40]{K} according to our measurement. The back-gate control does not shift the output voltage of such a BGR circuit as the output voltage of this type of reference is not a function of the threshold voltage of the MOS transistors. For different types of BJTs (lateral PNP and vertical NPN in this process) and different versions of transistor length (long channel and short channel) the output voltage characteristics along with the temperature are similar. The measurement results are presented in Fig. \ref{fig:BGR_BJT_FDSOI}. Although this type of BGR circuits will not function at \unit[4]{K}, they can still be used for higher temperature applications.

\section{Conclusions}

In conclusion, we have presented cryogenic operation of bandgap reference circuits fabricated in both \unit[0.35]{$\mu$m} SiGe BiCMOS and \unit[28]{nm} FDSOI CMOS processes. These references demonstrate low power operation across a wide-range of temperatures from \unit[300]{K} to \unit[4]{K}.  The measured power consumption of SiGe BGRs are \unit[1.5]{$\mu$W} and \unit[1.8]{$\mu$W} respectively at the measurement temperature of \unit[6]{K}. The reference circuits based on MOS-only topologies in \unit[28]{nm} FDSOI CMOS process remain functional on an SoC down to \unit[4]{K} temperature. Taken collectively these results and design techniques establish the performance of bandgap references for cryogenic applications, particularly those that relate to engineering a scalable control interface of a quantum computer.  

\section{Acknowledgments}
We thank R. Rouse for many useful discussions and help with managing tape-out of our \unit[28]{nm} circuits. This research was supported by Microsoft Corporation and the ARC Centre of Excellence for Engineered Quantum Systems (EQUS, CE170100009).

\bibliographystyle{IEEEtran}
\bibliography{IEEEabrv,sonic}

\vfill

\clearpage

\end{document}